\documentclass[conference]{IEEEtran}
\IEEEoverridecommandlockouts
\usepackage{cite}
\usepackage{amsmath,amssymb,amsfonts}
\usepackage{graphicx}
\usepackage{subcaption} 
\usepackage{textcomp}
\usepackage{xcolor}
\usepackage{comment}
\usepackage{multirow}
\usepackage{multicol}
\usepackage{booktabs}
\usepackage{colortbl}
\usepackage{dsfont}
\usepackage{dutchcal}
\usepackage{pifont}
\usepackage{algorithm}  
\usepackage{algpseudocode} 
\definecolor{lightgray}{gray}{0.9}
\definecolor{lightblue}{rgb}{0.68,0.85,0.9}

\def\BibTeX{{\rm B\kern-.05em{\sc i\kern-.025em b}\kern-.08em
    T\kern-.1667em\lower.7ex\hbox{E}\kern-.125emX}}
\begin{document}

\vspace{-15pt}
\title{Heterogeneity-Aware Memory Efficient Federated Learning via Progressive Layer Freezing\\
}

\author{{Yebo Wu\textsuperscript{1}}, {Li Li\textsuperscript{1,*}}, {Chunlin Tian\textsuperscript{1}}, {Tao Chang\textsuperscript{2}}, {Chi Lin\textsuperscript{3}}, {Cong Wang\textsuperscript{4}}, {Cheng-Zhong Xu\textsuperscript{1}} \\
\textsuperscript{1}State Key Lab of IoTSC, University of Macau \\
\textsuperscript{2}National University of Defense Technology, China\\
\textsuperscript{3}Dalian University of Technology, China\\
\textsuperscript{4}Zhejiang University, China\\
% \textsuperscript{*}Corresponding author\\
\thanks{\textsuperscript{*}Corresponding author. Email: llili@um.edu.mo}\\
}
\maketitle
\vspace{-15pt}
\begin{abstract}
Federated Learning (FL) emerges as a new learning paradigm that enables multiple devices to collaboratively train a shared model while preserving data privacy. However, intensive memory footprint during the training process severely bottlenecks the deployment of FL on resource-limited mobile devices in real-world cases. Thus, a framework that can effectively reduce the memory footprint while guaranteeing training efficiency and model accuracy is crucial for FL. 

In this paper, we propose SmartFreeze, a framework that effectively reduces the memory footprint by conducting the training in a progressive manner. Instead of updating the full model in each training round, SmartFreeze divides the shared model into blocks consisting of a specified number of layers. It first trains the front block with a well-designed output module, safely freezes it after convergence, and then triggers the training of the next one. This process iterates until the whole model has been successfully trained. In this way, the backward computation of the frozen blocks and the corresponding memory space for storing the intermediate outputs and gradients are effectively saved. Except for the progressive training framework, SmartFreeze consists of the following two core components: a pace controller and a participant selector. The pace controller is designed to effectively monitor the training progress of each block at runtime and safely freezes them after convergence while the participant selector selects the \emph{right} devices to participate in the training for each block by jointly considering the memory capacity, the statistical and system heterogeneity. Extensive experiments are conducted to evaluate the effectiveness of SmartFreeze on both simulation and hardware testbeds. The results demonstrate that SmartFreeze effectively reduces average memory usage by up to 82\%. Moreover, it simultaneously improves the model accuracy by up to 83.1\% and accelerates the training process up to 2.02$\times$.

\end{abstract}

\begin{IEEEkeywords}
Federated Learning, On-Device Training, Heterogeneous Memory
\end{IEEEkeywords}

\vspace{-15pt}
\section{Introduction}

%%版本1
% Federated Learning (FL)~\cite{mcmahan2017communication} emerges as a new learning paradigm that enables multiple mobile devices (e.g., smartphone and wearable device) to collaboratively train a shared model while preserving data privacy. Despite the promising benefits, one critical obstacle that prevents FL from widely supporting applications in real-world scenarios is the intensive memory footprint. In order to obtain high-quality analysis, the recently developed deep learning models are becoming deeper and wider. Meanwhile, larger memory space is required to store the parameters, intermediate outputs, and gradients during the training process. For instance, when training GPT-2~\cite{radford2019language}, even with a batch size of 1, the training process consumes over 20 GB of memory. However, the available RAM for existing mobile devices is quite limited, only ranging from 4 to 16 GB~\cite{memory_ram} for mainstream smartphones. The ever-increasing model complexity coupled with limited memory resources, unfortunately excludes low-end devices that would otherwise make unique contributions with their own data to the shared model. 

Federated Learning (FL)~\cite{mcmahan2017communication} emerges as a new learning paradigm that enables multiple devices to train a shared model while preserving data privacy collaboratively. Despite the promising benefits, the intensive memory footprint within the training procedure is a critical obstacle preventing FL from being effectively deployed in real-world scenarios. To retrieve high-quality analysis, the recently developed deep learning models are becoming deeper and wider. Meanwhile, it necessitates a larger memory space to accommodate parameters, intermediate outputs, and gradients during the training process. For instance, training ResNet50~\cite{he2016deep} on ImageNet with a batch size of 128 consumes over 26 GB of memory. Unfortunately, mainstream smartphones typically offer limited RAM, ranging from 4 to 16 GB~\cite{memory_ram}. Thus, the training process cannot be well supported. On the other hand, simply training the small models that can fit the existing devices severely deteriorates the representation capability and limits the application scope of FL.

\textbf{Limitation of Prior Arts.} Memory optimization for model training has been widely studied in centralized learning. Techniques such as gradient checkpointing~\cite{wang2022melon}, micro-batching~\cite{gim2022memory}, model compression~\cite{cao2022framework}, and host-device memory virtualization~\cite{huang2020swapadvisor} have been widely utilized to reduce memory footprint on the server side. However, the existing techniques cannot be directly used in the setting of FL. For instance, host-device memory virtualization aims to expand the available memory budget using host-side memory. Nonetheless, the CPU and GPU share the same physical memory and compete for concurrent data access on mobile SoCs. Micro-batching breaks down the large batch but slows down the training efficiency at the same time.  Model compression leads to prominent accuracy degradation.  The accuracy loss aggravates in FL through model aggregation.

%%版本1
% Recently, several works have proposed to tackle the issue of memory limitation in Federated Learning which can be mainly divided into the following two categories: 1) model heterogeneous training~\cite{li2019fedmd}~\cite{zhang2022fedzkt} and 2) partial training~\cite{diao2020heterofl}~\cite{yang2022partial}. The first category of approaches customizes models based on each client's memory capacity. They aggregate client models with various architectures through knowledge distillation~\cite{hinton2015distilling}. However, a large public dataset is imperative in this process which is often difficult to obtain in real-world settings. For partial training, subsets of the global model with different complexity are assigned to the participating clients according to their resource constraints. However, these approaches substantially compromise the model architecture and the model performance is deteriorated~\cite{kim2022depthfl}. Thus, a framework that can effectively break the memory wall while taking into account the training efficiency and model accuracy in real-world settings is crucial for Federated Learning. 

Recently, several approaches have been developed to tackle the resource limitation in FL, which can be mainly divided into the following two categories: 1) model heterogeneous learning~\cite{li2019fedmd}~\cite{zhang2022fedzkt}, and 2) partial training~\cite{diao2020heterofl}~\cite{yang2022partial}. For model heterogeneous learning, models are customized based on the memory capacity of each client, and those models with different architectures are aggregated through knowledge distillation~\cite{hinton2015distilling}. However, these approaches often require access to a public dataset for knowledge transfer, which is usually hard to obtain in real-world scenarios. Partial training assigns subsets of the global model with varying complexities to participating clients based on resource limitations. Nonetheless, these strategies severely compromise the model architecture and lead to a serious deterioration in model performance~\cite{kim2022depthfl}. Consequently, a framework that can effectively overcome memory limitations while balancing training efficiency and model accuracy in real-world FL scenarios is urgently required.

%%版本1
% \textbf{Observation and Challenge.} The following key observation guides us to break the memory wall of heterogeneous FL from a new perspective. The training progress of internal DNN layers differs significantly, and front layers can become well-trained much earlier than deep layers. This is because DNN features transition from being task-agnostic to task-specific from the first to the last layer. Thus, the front layers of a DNN often converge quickly, while the deep layers take a much longer time to train. Thus, training the model in a progressive manner can be promising. We can train the front layers first and safely freeze them after they are converged. Then the training of the latter part can be triggered. In this process, the backpropagation process and the corresponding memory space for the intermediate outputs, and gradients of the freezing layers can be effectively saved. Thus, peak memory usage is prominently reduced during the overall training procedure. However, designing such a progressive learning framework for FL is challenging and faces the following critical challenges. First, how to accurately capture the convergence state of layers? Second, in such a progressive learning framework, system and data heterogeneity are dynamically changing. How can client selection be effectively conducted in this scenario?

\textbf{Our Design.} The following key observation guides us to break the memory wall of heterogeneous FL from a new perspective. In FL, different layers of the shared model represent various convergence rates. The front layers stabilize much faster than the deep layers. This is for the reason that the front layers usually retrieve task-independent and relatively simple features while the latter layers capture the task-specific features. Thus, updating the full model during the whole training process leads to unnecessary computing resources. Progressively training the shared model can be promising. In this paper, we propose SmartFreeze, a framework that breaks the memory wall of FL from a new perspective. Unlike the vanilla FL that keeps updating the full model during the whole learning process, SmartFreeze conducts the training in a well-designed progressive manner. We can divide the model to be trained into blocks consisting of a specified number of layers corresponding to different stages, initiate the training with the front block, safely freeze it after convergence, and then iteratively trigger the training of the following blocks. In this way, the backpropagation process and the corresponding memory space for the intermediate activations and gradients can be effectively saved.

\textbf{Challenges and Techniques.} However, designing such a new learning paradigm is not trivial and it mainly faces the following challenges.

% However, designing such a new learning paradigm faces the following challenges.

\begin{itemize}
    \item \textbf{Construct end-to-end training for each block.} The end-to-end training of a block except the last one cannot be established without an output module. However, simply using a fully connected layer as the output module can severely disrupt the feature representation the block should capture in the original model. To address this challenge, we utilize convolutional layers to replace the blocks following the block being trained, thereby preserving the position information of the block being trained, and assisting the block in learning the expected feature representation.
    
    \item  \textbf{Safely perform layer freezing.} Inaccurate layer freezing will damage the model's performance. To accurately assess the training progress of each block, we design a pace controller that proposes a scalar-based block perturbation metric, which does not require access to raw training data.

    \item  \textbf{Select the \emph{right} devices for training each block.} As data and system heterogeneity can highly impact the convergence and training efficiency of a specific block. We design a participant selector that devises a graph-based community detection algorithm to select the \emph{right} devices to participate in the training by jointly taking into account memory capacity, data distribution, and runtime training capability.
\end{itemize}

% Except for the new learning paradigm, SmartFreeze consists of the following two core components: 1) pace controller and 2) participant selector. 
% Specifically, SmartFreeze first divides the shared model to be trained into multiple blocks and carefully designs output modules for all blocks except the last one to assist each block in learning expected feature representations. As the data and system heterogeneity across different devices severely impacts the convergence and training efficiency of a specific block, the participant selector designs a graph algorithm based on community detection to select the \emph{right} devices to participate in the training by jointly taking into account memory capacity, data distribution, and runtime training capability. Once the central server completes model aggregation in each round, the pace controller well evaluates the training progress of the current block from the scalar perspective, safely freezes it once converged, and iteratively triggers the training of the following blocks. In this way, the backward propagation of the frozen blocks and the memory space for the corresponding intermediate activations and gradients can be well saved. The whole process iterates until all the blocks have been successfully trained. 

\textbf{Evaluation.} Through extensive experiments, SmartFreeze effectively reduces the average memory footprint by up to 82\%, improves the model accuracy by up to 83.1\%, and simultaneously accelerates the training progress up to 2.02$\times$. To the best of our knowledge, SmartFreeze is the \emph{first} work that systematically tackles the memory limitation in FL while simultaneously taking into the model performance and training efficiency. The main contributions of this paper are summarized as follows: 

\begin{itemize}
    \item We propose SmartFreeze, an FL framework that effectively breaks the memory wall through progressive layer freezing while guaranteeing the model performance and training efficiency at the same time. 
    
    \item We design three core components, namely the method for constructing output modules, the pace controller, and the participant selector.
    
    % We first design an FL-based progressive training paradigm. Additionally, a method for constructing output modules to achieve end-to-end training for each block is proposed. Furthermore, a pace controller is designed to control the training pace of each block, and a participant selector is designed to select the \emph{right} devices to participate in the training.

    \item Extensive experiments are conducted to evaluate the effectiveness of SmartFreeze on both simulation and real-world devices with representative models and datasets. 
    
\end{itemize}

\vspace{-5pt}
\section{Motivation and Observation }

In this section, we try to answer the following questions to derive critical design principles for a memory-efficient FL system. 

\noindent{\textbf{Q1: Whether the existing memory reduction techniques designed for centralized learning can be directly applied in FL?}}

\begin{figure}[h]
\vspace{-5pt}
    \centering
        \begin{subfigure}{0.23\textwidth}
        \centering
        \includegraphics[width=\linewidth]{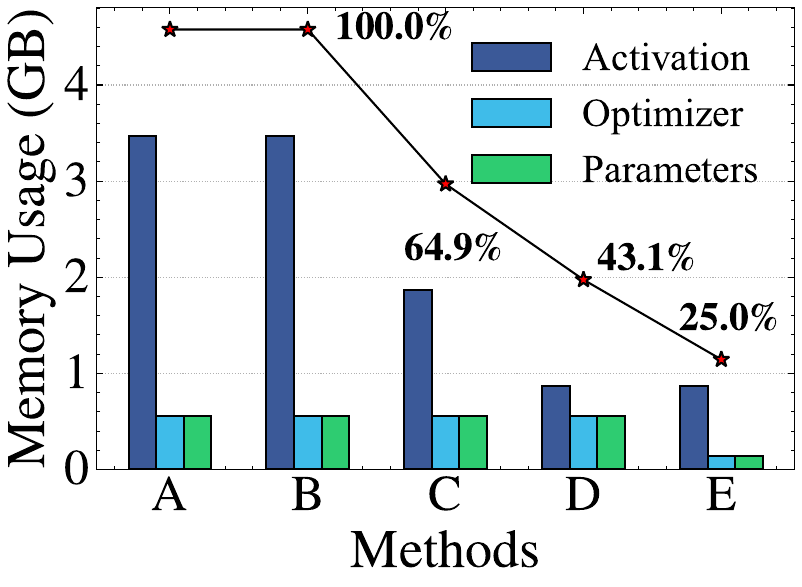}
        \caption{Memory Usage.}
        \label{sys1}
    \end{subfigure}    
    \begin{subfigure}{0.23\textwidth}
        \centering
        \includegraphics[width=\linewidth]{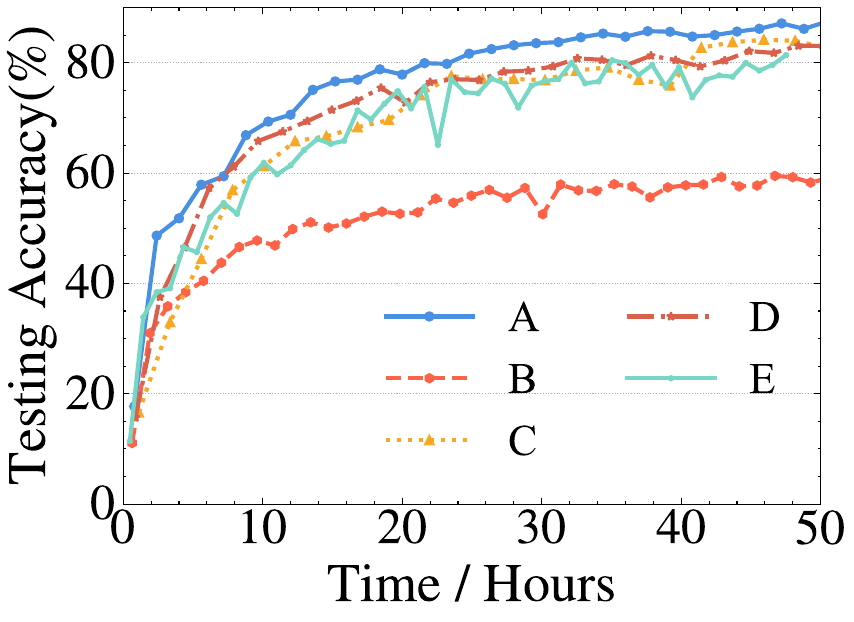}
        \caption{Accuracy $\&$ Training Time.}
        \label{sys12}
    \end{subfigure}\hfill
    \caption{The trade-off between accuracy and training overhead of various system-level memory optimization methods applied in FL. \textbf{A}: vanilla FL, \textbf{B}: local training with a single device, \textbf{C}: gradient checkpointing, \textbf{D}: gradient accumulation, \textbf{E}: model compression (low-precision training, int 8).}
    \label{sys_level}
    \vspace{-8pt}
\end{figure}

To investigate the effectiveness of existing memory reduction techniques, we establish the following experiment environment. Specifically, we use VGG16 as the global model and distribute the Tiny-ImageNet dataset to 100 devices, randomly selecting 10 devices in each training round. The representative strategies including: 1) gradient checkpointing, 2) gradient accumulation and 3) model compression are directly applied to the local training on each device. Concurrently, we assess each approach from three perspectives: 1) memory reduction, 2) model precision, and 3) training efficiency. Fig.~\ref{sys1} provides an overview of the memory reduction for each approach, where the X-axis represents distinct algorithms and memory components, while the Y-axis illustrates memory usage. Fig.~\ref{sys12} portrays time-accuracy curves, with the X-axis indicating local training time and the Y-axis denoting testing accuracy.

% To reduce memory usage within the training process, several existing techniques have been proposed. In this section, we assess their suitability for supporting FL. To simulate a more realistic FL scenario, we utilize the ImageNet dataset and distribute it among 100 devices. In each round, 10 devices are selected for training, with the global model based on VGG16 and a training batch size of 32. We directly apply the following techniques to the local training on each device. Simultaneously, we evaluate each scheme from three perspectives: 1) memory reduction, 2) model accuracy, and 3) training efficiency. Fig.~\ref{sys1} presents the memory usage for each algorithm, where the X-axis represents different algorithms and memory components, and the Y-axis represents memory usage. Fig.~\ref{sys12} illustrates the time-accuracy curves, with the X-axis representing local training time and the Y-axis representing accuracy.

\textit{Gradient checkpointing}~\cite{wang2022melon} involves the selective preservation of a subset of activations in memory while re-computing non-stored activations as needed during the backpropagation process to calculate gradients. This approach reduces memory usage to 64.9\%. However, it simultaneously increases local computation overhead, leading to an approximately 1.4$\times$ rise in training time. 

% \textit{Micro-batching}~\cite{gim2022memory}
% effectively reduces memory usage by minimizing batch sizes, thereby reducing the activation volume. However, it may lead to inaccurate global gradient estimates during the descent phase.

\textit{Gradient accumulation}~\cite{huang2019gpipe} represents an advanced version of micro-batching, collecting gradients from consecutive small batches in a dedicated buffer. While this approach reduces memory usage to 43.1\%, it also results in a 4\% decline in accuracy and a 1.1$\times$ increase in training time. This is for the reason that the gradients accumulated across multiple mini-batches may have different directions and magnitudes, leading to situations where gradient values are too small or too large during the accumulation process.

\textit{Model compression} aims to alleviate memory constraints by compressing the model's representations or structure. It encompasses techniques like low-precision training~\cite{hubara2017quantized} and knowledge distillation~\cite{hinton2015distilling}. Taking low-precision training (int 8) as an example, memory is reduced to 25\%. However, this reduction in memory is accompanied by a 12\% drop in accuracy. Regarding \textit{host-device memory virtualization}, the inherent shared memory design of mobile SoCs and other resource constraints make applying this technology a complex task in a mobile environment. In conclusion, existing system-level memory optimization techniques all come at the expense of either training accuracy or efficiency and are not directly applicable to FL.

\textit{$\#$Principle 1: Memory footprint, model accuracy and training efficiency should be jointly considered at the same time in a memory-efficient FL system. }

\noindent{\textbf{Q2: Whether different layers in a neural network have various convergence rates during the training process in FL?}}

\begin{figure}[h]
    \vspace{-5pt}
    \centering
    \begin{subfigure}{0.24\textwidth}
        \centering
        \includegraphics[width=\linewidth]{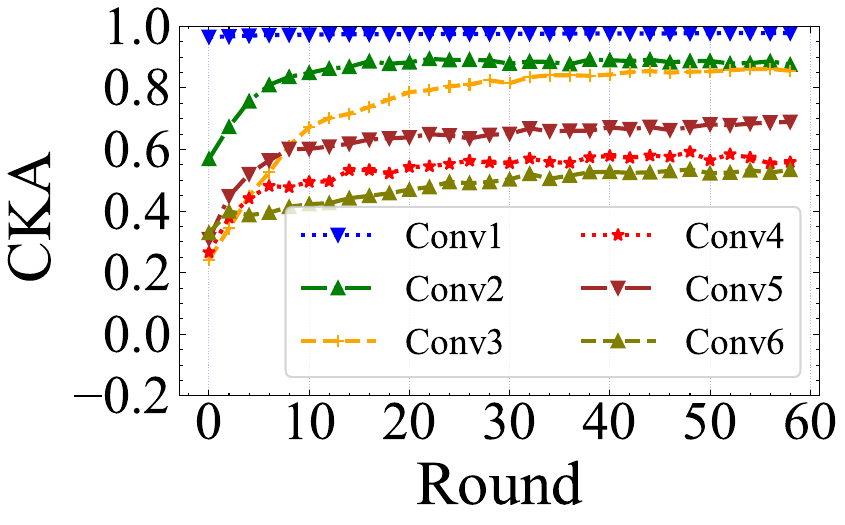}
        \caption{CIFAR10 (IID).}
        \label{cka1}
    \end{subfigure}\hfill
    \begin{subfigure}{0.24\textwidth}
        \centering
        \includegraphics[width=\linewidth]{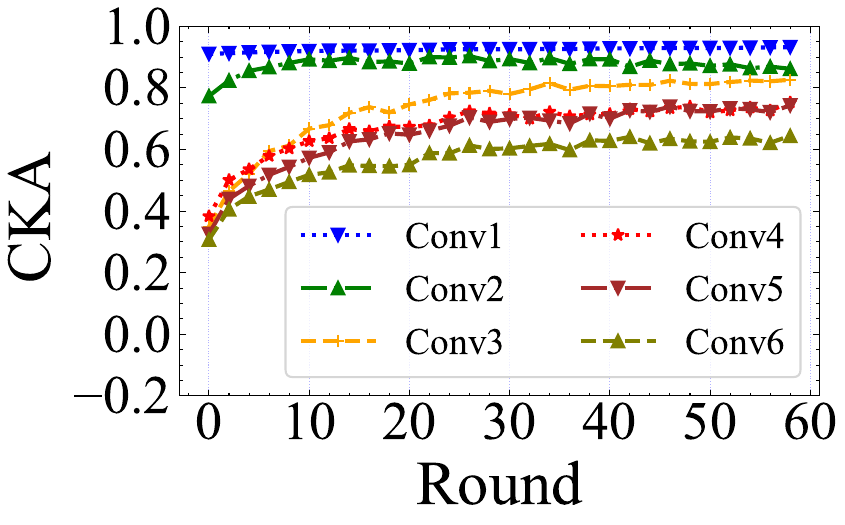}
        \caption{CIFAR10 (Non-IID).}
        \label{cka2}
    \end{subfigure}
    \caption{CKA of different layers in FL with VGG16.}
    \label{CKA}
    \vspace{-10pt}
\end{figure}

% While layer freezing can indeed accelerate model training, it's crucial to note that inappropriate layer-freezing strategies can potentially compromise the overall performance of the model. As the training process advances, the layers of the model undergo continuous updates, resulting in improved feature extraction capabilities. The convergence of a layer indicates a similarity in its feature extraction capacity to that of a fully trained layer.

% Layer freezing can indeed expedite the model training process. However, it is imperative to emphasize that inappropriate layer-freezing strategies have the potential to detrimentally affect the overall model performance. 

% As the training process progresses, the model's layers undergo continuous updates, leading to enhanced feature extraction capabilities. The representational similarity between the active training layer and its corresponding well-trained counterpart increases.
To investigate the convergence rates across different layers in FL, we conduct the following experiment. First, a VGG16 is trained on the entire CIFAR10 dataset, and we use the converged model as a reference model. The CIFAR10 dataset is then evenly distributed to 100 clients in the form of both independent and identically distributed (IID) and non-independent and identically distributed (Non-IID) respectively. After that, we train another VGG16 model using FL, with 20 clients randomly selected in each training round. 
The Centered Kernel Alignment (CKA)~\cite{kornblith2019similarity} between the activations of the shared model and the reference model is retrieved in each training round. Specifically, CKA quantifies the similarity between the representations of two layers and is computed by comparing the similarity of their respective outputs. A higher CKA signifies a stronger resemblance in feature extraction capabilities between the two layers. 

% In this case, a higher CKA means the corresponding layer is closer to convergence. 

\begin{figure}[t]
    \vspace{-5pt}
    \centering
    \begin{subfigure}{0.24\textwidth}
        \centering
        \includegraphics[width=\linewidth]{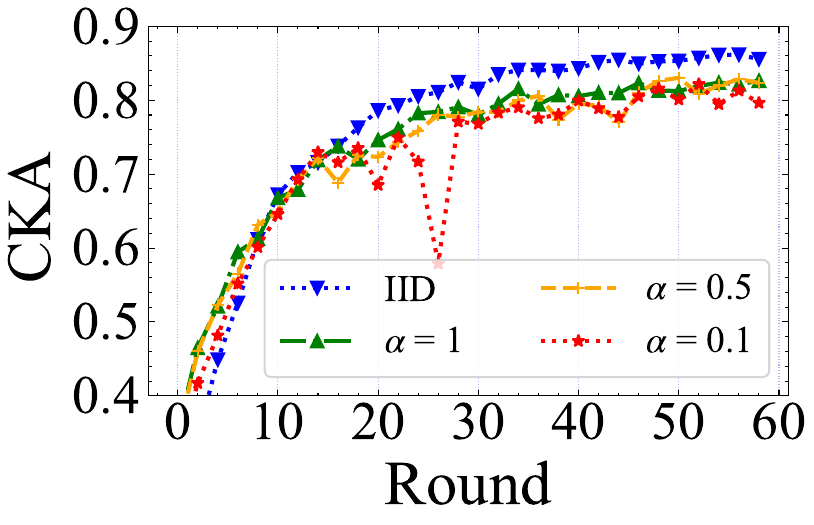}
        \caption{Conv3.}
        \label{degree1}
    \end{subfigure}\hfill
    \begin{subfigure}{0.24\textwidth}
        \centering
        \includegraphics[width=\linewidth]{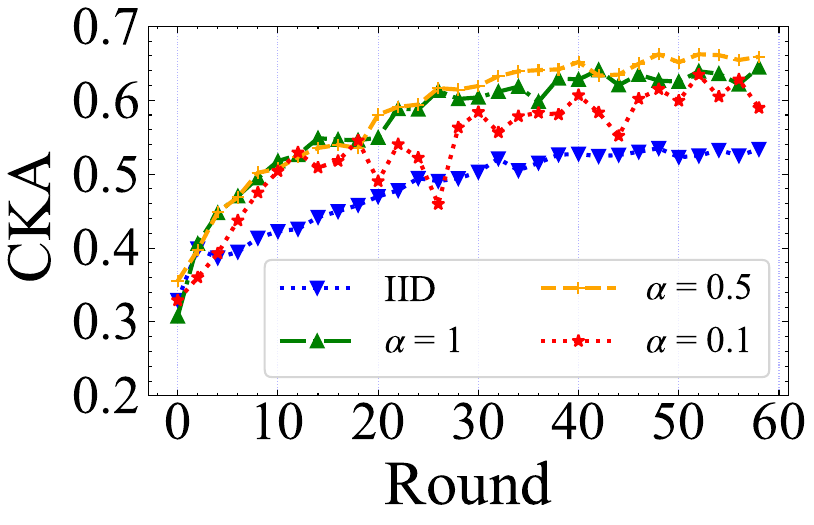}
        \caption{Conv6.}
        \label{degree2}
    \end{subfigure}
    \caption{CKA with different degrees of Non-IID in FL.} 
    
    % $\alpha$ represents the Non-IID degree of data distribution~\cite{xu2022fedcorr}. $Conv3$/$Conv6$ represents the third/sixth convolutional layer.}
    \label{degree}
    \vspace{-20pt}
\end{figure}

Fig.~\ref{CKA} illustrates the results of CKA for the first six convolutional layers. From Fig.~\ref{cka1}, it can be observed that the CKA values of the six layers continuously increase as training proceeds, eventually stabilizing around the 1st, 15th, 35th, 40th, 45th, and 50th rounds respectively. Various layers have totally different convergence rates. The front layers quickly converge, while the later layers exhibit a slow convergence rate. This is for the reason that the front layers extract task-independent and simple features while the later layers capture task-specific and intricate features~\cite{chen2022layer}. Thus, updating all the layers from the beginning to the end of the training procedure can cause unnecessary computation overhead, providing an opportunity for layer freezing. This phenomenon is more pronounced under the Non-IID setting, as illustrated in Fig.~\ref{cka2}. This is because the global data distribution, which means the distribution of total data involved in devices participating in each training round, is Non-IID, resulting in biases in both the direction and magnitude of parameter updates. Additionally, Fig.~\ref{degree} displays the CKA curves of the third and sixth convolutional layers at different degrees of Non-IID. Here, $\alpha$ represents the degree of Non-IID, where smaller $\alpha$ values indicate higher levels of Non-IID. From Fig.~\ref{degree}, it can be seen that compared to the IID setting, the higher the degree of Non-IID, the slower the convergence speed of the layers, especially for the later layers. This is because as the degree of Non-IID increases, the degree of Non-IID of the global data distribution involved in each training round also increases.

\textit{$\#$Principle 2: Different layers have various convergence rates during the training process. The data distribution across different clients prominently impacts the convergence rate for each layer. Selecting clients to ensure that the overall distribution of data involved in each training round is IID can accelerate convergence.}

\vspace{-5pt}
\section{SmartFreeze Overview}

Motivated by the above observations, we propose SmartFreeze, a new FL framework that effectively reduces the memory footprint while effectively balancing the model accuracy and training efficiency at the same time.

% Instead of updating the full model in each training round, SmartFreeze conducts the model training in a progressive manner. It divides the shared model into multiple blocks and separate the whole training process into corresponding stages. The first block is trained at the starting point. We safely freeze it after convergence and concatenate the next block to enter the second training stage. The training proceeds until  all the blocks have been fully trained. Thus, the memory footprint for the intermediate activations of the frozen blocks is effectively saved.

\begin{figure}[!t]
  \centering
  \includegraphics[width=0.9\linewidth]{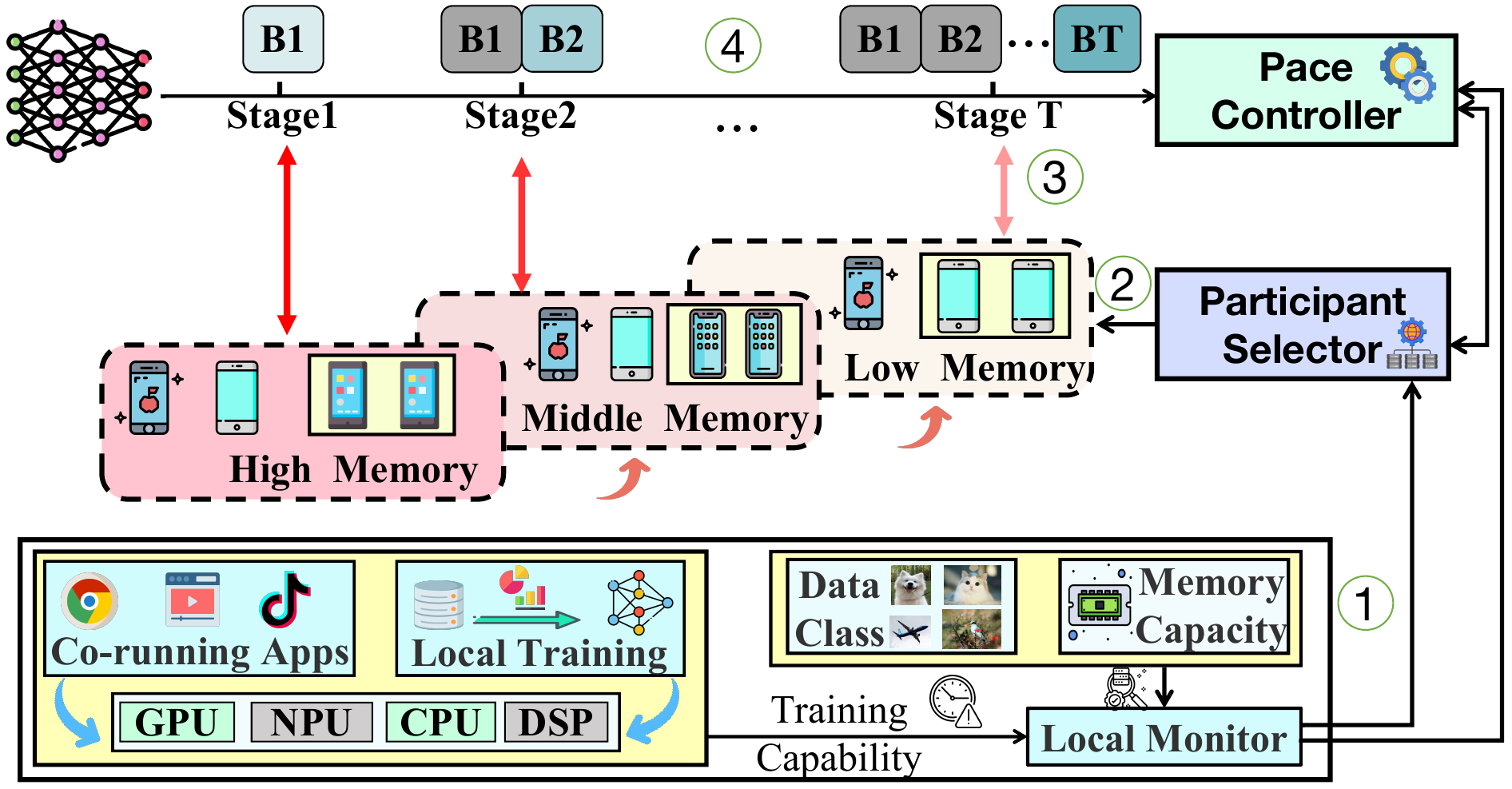}
  \caption{Architecture and workflow of SmartFreeze.
  The Pace Controller and Participant Selector are deployed on the server side. Local Monitor is deployed on each device. }
  \label{Fig7}
  \vspace{-18pt}
\end{figure}

SmartFreeze mainly consists of the following components. The participant selector, hosted on the server side, intelligently selects the \emph{right} devices to participate in the training for each block by jointly considering the system, data heterogeneity, and memory capacity. The Pace controller, deployed on the server side, determines when a block can be safely frozen and the next training stage can be triggered. A lightweight local monitor is deployed on the device side to obtain the runtime training capabilities, data importance of each round, gradient vectors, and memory resources. Fig.~\ref{Fig7} represents the architecture and workflow of SmartFreeze. It can be mainly divided into the following steps:

\begin{enumerate}
    \item\label{enu:mr} System Initialization (\ding{172} in Fig.~\ref{Fig7}). At the initialization step, SmartFreeze divides the shared model to be trained into $T$ blocks. Each block contains a specific number of layers which can be configured offline. The local monitor deployed on each device retrieves and sends the following information to the central server: 1) memory capacity, 2) the gradient vectors of the global model's output layer (only once), 3) runtime training capability, and 4) local data loss.
    
    % Then, the Participant selector groups the clients based on the distribution relationship of the clients obtained from gradient vector similarity.
    
    \item\label{enu:ms} Client Selection (\ding{173} in Fig.~\ref{Fig7}). According to the received information, the Participant Selector selects the \emph{right} devices for the first block according to their memory capacity, training capability, data distribution, and local data loss to balance the training performance and progress well. Then, it broadcasts the first block of the shared model to the selected devices and triggers the first training stage.

    \item\label{enu:mt} Stage-based Training (\ding{174} in Fig.~\ref{Fig7}). The selected devices then conduct the local training in each training round and send the updated block to the central server for aggregation, along with their local losses. This process iterates until the pace controller determines that the first block has converged. 

    \item\label{enu:mu} Model Growth (\ding{175} in Fig.~\ref{Fig7}). After that, SmartFreeze freezes the first block and concatenates the second block. Step \ref{enu:mr} to step \ref{enu:mu} iterates until the last block completes the training procedure.

    % In this step, the sub-model of the current stage is collaboratively optimized by the clients who have enough memory to support the training. In each round, the coordinator distributes the current model to the best client set $S$, which is weighted and selected from the groups organized by the community detector. The weight is measured based on the training loss and training time when training the sub-model, and for a newly incoming client, it is set to infinity. The selected clients engage in local training and send the updated parameters along with their local losses to the central server. At last, the central server aggregates the received local models into a new global model.

    % In each round, the training coordinator filters the client according to the runtime training capability collected by the local monitor. Then, the coordinator distributes the current model to the best client set $S$, which is weighted and selected from the groups organized by the community detector. The weight is the data importance measured by the training loss on training the sub-model, and it is set to infinite for a newly incoming client. The selected clients engage in local training and send the updated parameters along with their local losses to the central server. At last, the central server aggregates the received local models into a new global model.   
\end{enumerate}

\begin{comment}

It's important to highlight that the communication overhead associated with data transmitted by the local monitor is exceptionally low. 
Specifically, we only need to transmit the output layer gradient vector once, and the communication overhead related to running status and available memory resources, among other details, is negligible. 
Moreover, utilizing the output layer gradient alone for client similarity inference incurs minimal additional computational overhead, all while preserving privacy and without altering the model aggregation process. 
Consequently, SmartFreeze can seamlessly integrate with algorithms like SecAgg~\cite{bonawitz2017practical} and Differential Privacy (DP)~\cite{dwork2014algorithmic} with minimal adjustments to bolster its privacy protection capabilities. 

\end{comment}

% To sum up, as training progresses, more clients can gradually become involved in SmartFreeze, thereby increasing the participation rate and addressing the following challenges: (1) By having clients synchronously train the same model parameters at each stage, the issue of parameter mismatch is resolved; (2) Progressive model growth ensures the attainment of the desired global model over time; (3) This allows more clients to participate, effectively utilizing the data information from low-memory devices.
\vspace{-6.5pt}
\section{System Design}

In this section, we discuss the detailed design of SmartFreeze. Specifically, we aim to clearly illustrate the following three parts: 1) the overall progressive training paradigm, 2) pace control across different stages, and 3) participant selection for each stage.

% In this section, we commence by elucidating the procedure of progressive training with layer freezing. Subsequently, we describe how to use the model scheduler to control the pacing of different stages. Finally, we elaborate on the client selection process for each stage by leveraging the community detector, stage-based memory, time, and data model.

\vspace{-5pt}
\subsection{Progressive Training with Layer Freezing}\label{section4_1}

\begin{figure}[!t]
  \centering
  \includegraphics[width=0.9\linewidth]{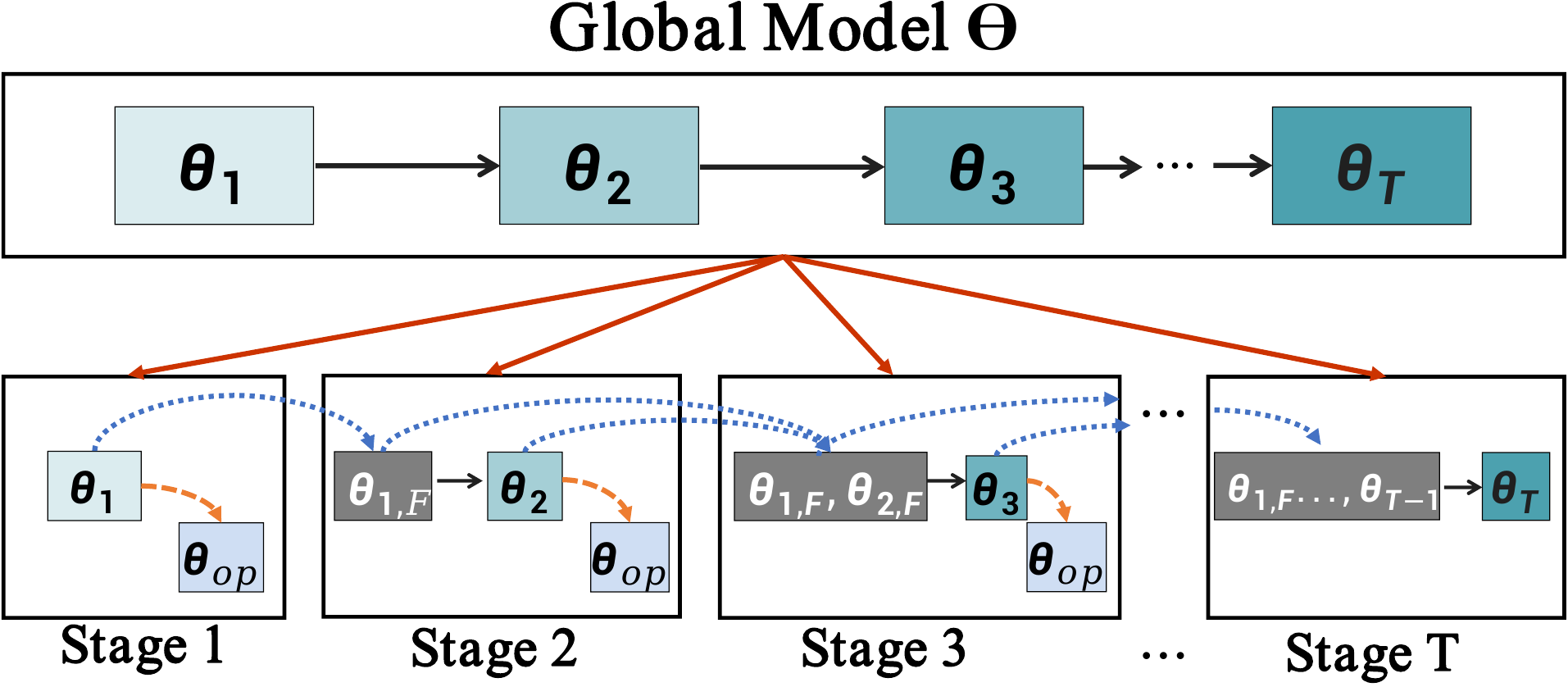}
   \caption{Progressive training with layer freezing. The global model $\Theta$ is split into $[\theta_1, \theta_2,..., \theta_{T}]$ according to the model's structure. In each stage $t$, an additional output module $\theta_{op}$ is attached to $\theta_t$ for training. After training is completed, it will be frozen and proceed to next stage. Iterative, the training part of the model gradually grows until it reaches the target model.}
  \label{modelg}
  \vspace{-18pt}
\end{figure}

Fig.~\ref{modelg} demonstrates the process of progressive training. We use $\theta_{t}$ to represent the currently training block and $\theta_{t,F}$ to represent the frozen block. For the global model $\Theta$, based on its original architecture and client memory distribution, 
we partition it into $T$ blocks ($\Theta = [\theta_1, \theta_2, \cdots, \theta_T]$). Each block may consist of multiple layers to preserve the inherent atom architecture, such as a residual block. Specifically, the sub-model trained in each stage can be represented as $\Theta_{t}$  ($\Theta_{t} = [\theta_{1,F}, \theta_{2,F}, \cdots, \theta_{t-1,F},\theta_t]$).

% where $\theta_t$ is the only block being trained in the current stage and $\theta_{1,F}$ - $\theta_{t-1,F}$ represent the frozen blocks.

% For a partial training based approach, SmartFreeze progressively trains $\Theta = [\theta_1, \theta_2, \cdots, \theta_T]$.

% To avoid destroying the atom architecture of the model, each of the $\theta_i$ is composed of several layers or functional block.

However, the sub-model $\Theta_t$, where $1 \leq t < T$, cannot finish the whole training procedure independently due to the lack of an output module. Therefore, to address this issue, an output module $\theta_{op}$ is concatenated to the sub-model to ensure independent training, thus facilitating the completion of this stage-based progressive training process. Therefore, the sub-model of stage $t$ can be represented as $\Theta_{t}$ ($[\theta_{1,F}, \theta_{2,F}, \cdots, \theta_t, \theta_{op}]$).

In this stage-wise training process, the execution of each round mainly includes the following steps: 1) The central server determines the sub-model $\Theta_{t}^{r}$ ($[\theta_{1,F}, \theta_{2,F}, \cdots, \theta_t, \theta_{op}]$) to be trained in the current round $r$; 2) The central server sends $\Theta_{t}^{r}$ to the selected client set $S$; 3) Clients perform end-to-end training on local datasets. Due to changes in the computation graph, the forward propagation of local training passes through layers $[\theta_{1,F}, \theta_{2,F}, \cdots, \theta_t, \theta_{op}]$. In contrast, the backward propagation goes through layers $[\theta_{op}, \theta_{t}]$. After completing local training, we only upload the updated model parameters ($[\theta_{t,i}^{r}, \theta_{op,i}^{r}], \text{where~}
i \in S$); 4) The central server then aggregates the parameters using Eq.~\eqref{FL_process} and replaces the updated parameters at their respective positions.
\vspace{-3pt}
\begin{equation}
    \label{FL_process}
    [\theta_{t,g}^{r}, \theta_{op,g}^{r}] = \sum_{i \in S} \frac{|D_{i}|}{|D|} ([\theta_{t,i}^{r}, \theta_{op,i}^{r}])
    \vspace{-8pt}
\end{equation}
where $|D_{i}|$ signifies the local dataset size of client $i$, $|D|$ represents the total dataset size of participating clients, and $[\theta_{t,i}^{r}, \theta_{op,i}^{r}]$ stands for the locally updated model parameters on round $r$ by client $i$. After completing aggregation, the Pace Controller assesses the convergence of the current block and determines whether to freeze it and proceed to the next stage. The stage $t+1$ integrates the blocks in the preceding $t$ stages to perform model growth and conduct training for the current stage. After $T$ training stages, the trained model grows to the target model. 

The key challenge in progressive training is how to design the output module for each sub-model to be trained in each stage. Simply adding a fully connected layer as an output module would compromise the final model's performance. This is because the block trained at that stage is encouraged to learn high-level features most beneficial for classification to reduce the training loss. Nevertheless, it is well-known that each block in a model has its inherent role, with earlier blocks primarily responsible for extracting low-level features. Consequently, this can severely disrupt the feature representation the block should capture in the original model.

% this can result in each block's inability to learn the intended knowledge effectively.

In order to maintain the role of each block as much as possible, we design the output modules for each block based on their specific position. Specifically, for the block currently being trained, each subsequent block is replaced with a convolution layer to emulate its position in the original model. For instance, when training ResNet-18's first residual block (4 blocks in total), we append three convolutional layers and a fully connected layer as the output module behind it to preserve the first block's position information. The intuition behind this strategy is to mimic the original model architecture through this construction method, thereby preventing functional disarray.

\vspace{-3pt}
\subsection{Pace Controller}

The pace controller is designed to evaluate the training progress of each block at runtime and safely freezes them after convergence.

% The key challenge in designing the Pace Controller lies in accurately determining the convergence status of the current block to enable secure freezing. 

In centralized learning, a reference model is utilized for assessment~\cite{wang2023egeria}, but this method requires sending the raw training data to the reference model and the model being trained for comparison. However, in FL raw training data is not accessible to the central server for privacy preserving. Therefore, to accurately evaluate the convergence status of each block, we delve into a novel perspective to study the status of each scalar during the training process. In specific, we define the effective update of a scalar as $\frac{||\sum_{q=0}^{Q-1} U^{r-q}||}{\sum_{q=0}^{Q-1}||U^{r-q}||}$, where $U^{r} = w^{r}-w^{r-1}$ indicates the scalar's update at round $r$, $w$ denotes the value of the scalar, and $Q$ represents the observation window size. If a parameter converges, its effective update will tend to approach zero. The rationale for this phenomenon lies in the fact that when a scalar converges, these recent updates tend to consistently approach zero, or the conflicts between updates intensify, leading to their mutual cancellation and ultimately resulting in a reduction of effective update~\cite{chen2021communication}. We define the sum of effective updates of all scalars within each block as the block perturbation to reflect the learning status of each block, which is defined as follows:
\vspace{-4pt}
\begin{equation}
    P_{t}^{r,Q} = \frac{\left\| \sum_{q=0}^{Q-1} W_{t}^{r-q} \right\|}{\sum_{q=0}^{Q-1} \left\| W_{t}^{r-q} \right\|}
    \label{eq_layerP2}
    \vspace{-5pt}
\end{equation}
where $r$ represents the current round,  $W_{t}^{r} = \theta_{t}^{r}-\theta_{t}^{r-1}$, and $\theta_{t}$ represents the block $t$. In determining when to freeze the block trained at the current stage, we first calculate block perturbation using Eq.~\eqref{eq_layerP2}. Then, we employ a sliding window of size $H$ to smooth the block perturbation. The resulting smoothed block perturbation is expressed as:
\vspace{-3pt}
\begin{equation}
\overline{P_{t}^{r,Q}} = \begin{cases}
\frac{P_{t}^{r-H,Q} + \dots + P_{t}^{r,Q}}{H}, & \text{if} \, r \geq H \\
\frac{P_{t}^{0,Q} + \dots + P_{t}^{r,Q}}{r}, & \text{if} \, r < H \\
\end{cases}
\label{window_smooth}
\vspace{-5pt}
\end{equation}
where $P_{t}^{r,Q}$ represents the block perturbation for block $t$ in the round $r$. Then, we use linear least-squares regression to fit the recently obtained smoothed block perturbation to judge whether the curve converges. If the absolute value of the curve slop remains below the threshold $\varLambda$ for a consecutive $\mu$ times, the block is considered to have converged and can be frozen. For the choice of $H$ and $\mu$, if $H$ is larger, it results in smoother block perturbation, which can help mitigate extreme values. On the other hand, if $\mu$ is larger, it requires a longer waiting period for freezing. Therefore, the selection of $H$ and $\mu$ can be adjusted based on the block perturbation curve.

\vspace{-5pt}
\subsection{Participant Selector}

% In this section, we have designed a community detection algorithm to group clients based on the data distribution. Additionally, we establish stage-based memory, time, and data models. Taking into account data and system heterogeneity, we designed a stage-based heterogeneity-aware client selection algorithm.

According to \textit{$\#$Principle 2}, the data distribution prominently impacts the convergence rate of each block. At the same time, the memory capacity of a device determines the participating feasibility, while computing capability decides the training efficiency for each stage. Thus, to meet \textit{$\#$Principle 1}, SmartFreeze intelligently selects the participating clients for each training stage in order to jointly consider the model accuracy and training efficiency. In this section, we first introduce the models and problem formulation. After that, we present the detailed design of participant selection.

% In this section, we establish stage-based memory, time, and data models. Taking into account data and system heterogeneity, we designed a stage-based heterogeneity-aware client selection algorithm. To address this compound optimization problem, we have devised a community detection algorithm to decouple the problem.

\subsubsection{Stage-Based Memory Model}\label{subsection_memory_model}

When employing stage-based progressive training, the memory usage for each stage $\mathcal{T}$, can be represented as:
\vspace{-8pt}
\begin{equation}
\begin{split}
    M_{\mathcal{T}} = (\theta_{\mathcal{T}}^{a}+\theta_{op}^{a}) \cdot 2 + (\sum_{t=1}^{\mathcal{T}} |\theta_t| + |\theta_{op}|) + {M}_{{optimizer},\mathcal{T}} \\
    +\max([\theta_{1,F}^{a}, \theta_{2,F}^{a}, \ldots, \theta_{\mathcal{T}}^{a}, \theta_{op}^{a}])
\end{split}
\label{eq_memory2}
\vspace{-8pt}
\end{equation}
where $\theta_{\mathcal{T}}^{a}$ represents the activations generated within the block $\theta_{\mathcal{T}}$, and the reason for multiplying it by two lies in the necessity of computing gradients for the activations during the process of backward propagation. $|\theta_{t}|$ signifies the memory occupied by the block parameters, $M_{{optimizer},\mathcal{T}}$ represents the memory occupied by the optimizer, and the last term represents the maximum activation generated during the forward propagation. It's worth noting that for layers $[\theta_{1,F}, \theta_{2,F},..., \theta_{\mathcal{T-1},F}]$, which have been frozen, there's no need for backpropagation, and correspondingly, the activations don't need to be stored.

\begin{comment}

Comparing Equation.~\eqref{eq_memory2} and Equation.~\eqref{eq_memory1}, it is evident that in stage-wise training, both activation and optimizer terms are only related to the currently trained layer, and the parameters is only dependent on the current sub-model. Although an additional term for the maximum activation layer output during forward propagation is introduced, this is negligible in comparison to the reduction in activations. 

Furthermore, as shown in Fig.~\ref{Fig4}, it is evident that when freezing the preceding layers, a significant reduction in memory can be achieved. Thus, through this stage-wise progressive training paradigm, as stages progress, the memory requirements gradually decrease, enabling the involvement of more low-memory devices.

By using Equation.~\eqref{eq_memory2}, we can calculate the required memory for each stage. It helps Participant Selector identify the available client set for each stage.

\end{comment}

\subsubsection{Stage-Based System Time Model} 

% When employing stage-based progressive training, the local training time for client $i$ at stage $t$ can be represented as:

Taking into account the distinct computation graphs for different stages, we can represent floating-point operations (FLOPs) required for the current sub-model training at stage $\mathcal{T}$ as:
\vspace{-5pt}
\begin{equation}
    \label{flops}
    FLOPs_{\mathcal{T}} = (\sum_{t=1}^{\mathcal{T-1}} \theta_{t,F}^{FP}+ \theta_{\mathcal{T}}^{FP}+\theta_{op}^{FP}) + (\theta_{\mathcal{T}}^{BP}+\theta_{op}^{BP})
    \vspace{-5pt}
\end{equation}
where $\theta_{t}^{FP}$ represents the FLOPs for forward propagation and $\theta_{t}^{BP}$ represents the FLOPs for backward propagation for block $t$. Due to differences in computation workload at each stage, the time to complete local training on the same device varies. We represent the training completion time for client $i$ at stage $t$ as: 
\begin{equation}
    \textbf{t}_{t}^{i} = \frac{\rho  \cdot FLOPs_{t} \cdot |D_{i}|}{c_{i}}
    \vspace{-5pt}
\end{equation}
where $\rho$ represents the coefficient and $\rho \cdot \mathrm{FLOPs}_{t}$ collectively represents the number of instructions required to process a training data during stage $t$. We determine this value through offline analysis. $c_{i}$ represents the runtime training capability of the device. Since we employ a synchronous method for model aggregation at each round, the time of each round is determined by the client within the selected client set $S$ with the longest completion time. Therefore, the time for each round $r$ at stage $t$ can be expressed as:
\vspace{-3pt}
\begin{equation}
    \textbf{T}_{r}(S,t) = max_{i \in S}(\textbf{t}_{t}^{i})
        \vspace{-5pt}
\end{equation}

\subsubsection{Stage-Based Data Model}
\
\newline
\textbf{Stage-Based Data Diversity.} Data diversity is characterized by the richness of data samples participating in each training round, which serves to mitigate the impact of Non-IID. Especially in this progressive training paradigm, the number of clients participating in each stage is limited, and the data available within each stage's client set varies. Thus, selecting data of sufficient richness is of paramount importance. We define data diversity as $Div(S,t) = \frac{1}{\sum_{i,j \in S} \Omega_{(i,j)}}$, where $\Omega_{(i,j)}$ represents the similarity between clients $i$ and $j$. To accurately infer similarity between clients under memory constraints, we use Eq.~\eqref{similarity} for calculations.
\begin{equation}
    \Omega_{(i,j)} = \frac{\nabla f_{i}(w_{op}) \cdot \nabla f_{j}(w_{op})}{\sqrt{||\nabla f_{i}(w_{op})||_{2}^{2}}\sqrt{||\nabla f_{j}(w_{op})||_{2}^{2}}}
    \label{similarity}
\end{equation}
where $\nabla f_{i}(w_{op})$ represents the gradient vectors of the output layer when training only the global model's output layer. This is for the reason that, through our experiments, we discover that relying solely on the output layer to infer the data distribution relationships among clients yields more accurate results. Moreover, reducing local training epochs enhances precision. Using Eq.~\eqref{similarity}, we can obtain the client similarity matrix $\Omega$. Hence, to enrich the selected client data, we aim to maximize $Div(S, t)$ between the selected clients in each round at stage $t$.

\noindent\textbf{Stage-Based Data Importance.} The contribution of a client to a certain block during the training stage evolves as the training proceeds. On the one hand, some data may have been well-learned in previous stages. On the other hand, in each stage, new clients join FL, and the model has not yet learned their data. Thus, we define the data importance of client $i$ at stage $t$ as:
\vspace{-4pt}
\begin{equation}
    I_{t,i} = \sum_{d=1}^{|D_{i}|}L(\Theta_{t},(x_{d},y_{d}))
    \vspace{-5pt}
    \label{importance}
\end{equation}
where $L(\Theta_{t},(x_{d},y_{d}))$ represents the local training data loss and $(x_{d},y_{d})$ represents a training data sample. The rationale for employing Eq.~\eqref{importance} to quantify the data importance is that losses are naturally computed during training, incurring no additional computation overhead. At the beginning of each stage, we allow all clients eligible for the current stage to compute the data importance.

To sum up, based on the stage-based data diversity model and stage-based data importance model, we can define the stage-based data model as follows:
\vspace{-3pt}
\begin{equation}
   Div(S,t)+\sum_{i \in S}I_{t,i}
   \label{datamodel}
\end{equation}
\vspace{-3pt}

\begin{figure*}[!ht]  
    \centering
    \includegraphics[width=1\linewidth]{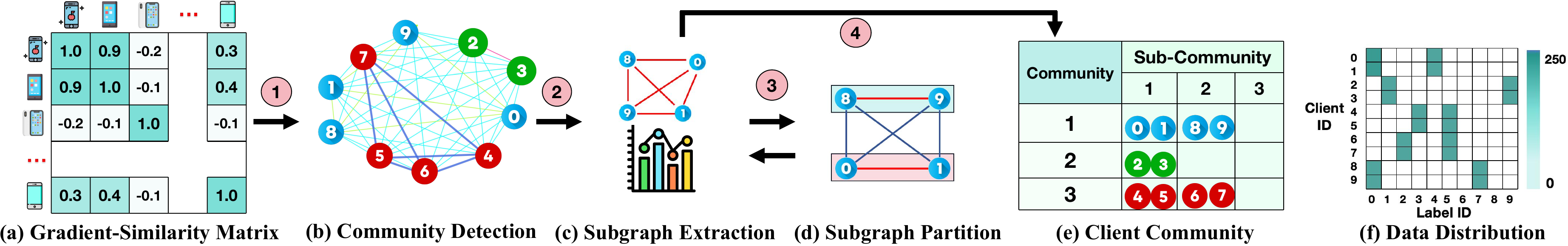}
    \caption{Procedure for community detection using the RL-CD algorithm. The process begins with the obtained similarity matrix (a), followed by an initial division using the Louvain algorithm (b). Subsequently, an evaluation of the weight distribution within the community (c) determines whether further community divisions are needed (d). For the final result (e), when we select clients, we should try to cover clients from each community/sub-community. }
    \label{Fig11}
    \vspace{-18pt}
\end{figure*}

\subsubsection{Problem Formulation}

Considering the stage-based memory, time, and data model, we formulate the client selection problem as a constrained optimization problem for each stage, specifically defined as follows (at stage $t$):
\vspace{-3pt}
\begin{align}
    &\Phi_{t} = \max\left({Div(S,t)+\sum_{i \in S}I_{t,i}} - \lambda \textbf{T}_{r}(S,t)\right)\label{Formulate17} \\
    s.t. \quad & M(i,t) \geq M_{\text{train}} (\Theta_{t}) \label{Formulate18} \\
    & \sum_{i=1}^{|N|} |D_{i}| \cdot B_{i} \geq \Gamma \label{Formulate19} \\
    & \sum_{i=1}^{|N|} \mathds{1} \geq \varphi, \mathds{1} =
    \begin{cases}
    1, &  M(i,t) \geq M_{\text{train}} (\Theta_{t}) \\
    0, & M(i,t) < M_{\text{train}}(\Theta_{t})
    \end{cases}
    \label{Formulate20}
\end{align}
where $1 \leq t \leq T$, $B_{i}$ serves as a binary variable indicating the selection status of client $i$, and $|N|$ represents the total number of clients. Additionally, we introduce the utility metric for client $i$, denoted as $Util_{i}$, calculated as $I_{t,i}-\lambda \textbf{t}_{t}^{i}$, which helps assess the client's suitability. Our objective is to determine the optimal client set $S$ for each round by optimizing the Eq.~\eqref{Formulate17}. Eq.~\eqref{Formulate18} requires the selected client $i$ for stage $t$ to possess adequate available memory to accommodate the training of $\Theta_{t}$. Eq.~\eqref{Formulate19} is to enhance the training efficiency for each round. Eq.~\eqref{Formulate20} is designed to ensure that in each stage, the number of participating clients exceeds the threshold $\varphi$ to guarantee the training effectiveness. The above problem is a compound optimization problem, and brute force solving is not feasible in FL. To overcome this challenge, we design a graph-based community detection algorithm, which untangles the intricacy of the problem, transforming it into an endeavour focused on maximizing clients $Util$.

% it consists of variants of the 0-1 knapsack problem~\cite{singh2011solving} and the linear bottleneck assignment problem (LBAP)~\cite{pferschy1997solution}. The naive solution is to list all combinations to meet the data diversity through brute force, and then estimate the difference between the calculation time of all combinations and their data importance, and the combination with the smallest result is the solution. 

% However, in the realm of cross-device FL, where the number of clients is substantial, the solution space expands exponentially with the growing client count~\cite{singh2011solving}. Consequently, the brute-force approach becomes unviable due to its escalating computational complexity. 

\subsubsection{Community Detection} 

To decouple Eq.~\eqref{Formulate17}, we design a graph-based community detection algorithm that divides clients into different communities based on the similarity matrix $\Omega$, aiming to maximize $Div(S,t)$.

The Louvain algorithm~\cite{blondel2008fast} is a type of community detection, but directly applying it results in too coarse-grained communities that cannot accurately capture the data differences among clients. Therefore, we design a robust Louvain community detection (RL-CD) algorithm in response to this challenge. RL-CD enhances community partitioning through iterative application of the Louvain algorithm, incorporating a weight distribution sharpening technique. During the partitioning process, we introduce the following criteria as conditions to halt the community division. $Standard_{stop}$: There is no clear hierarchy of weight distribution within the communities. Moreover, we give an example with 10 clients in Fig.~\ref{Fig11} to illustrate the process, mainly consisting of the following steps. The data distribution is as shown in Fig.~\ref{Fig11}f.

\textbf{Step 1}: Initially, we compute the similarity matrix $\Omega$ using Eq.~\eqref{similarity}, as shown in Fig.~\ref{Fig11}a. We then construct a graph using $\Omega$ and apply the Louvain algorithm for community detection. Fig.~\ref{Fig11}b indicates that although clients with similar data distributions are grouped, it does not precisely capture data distribution relationships. For example, clients 0, 1, 8, and 9 are placed in the same community because they all have label 0. However, from Fig.~\ref{Fig11}f, it can be seen that there are still significant differences among them. Our goal is to distinguish clients 0 and 1 from 8 and 9. 

\textbf{Step 2}: We extract subgraphs for each community from the previous partition and quantify the distribution of similarity weights using histograms. Fig.~\ref{Fig11}c displays a clear hierarchical weight distribution within the community, indicating that it does not meet $Standard_{stop}$ and requires further partitioning, proceeding to step 3. If further partitioning is not necessary, it directly proceeds to step 4.

\textbf{Step 3}: To facilitate further partitioning, we utilize the median as a threshold to sharpen the similarity weight and apply the Louvain algorithm to partition the sharpened weight distribution. From Fig.~\ref{Fig11}d, it's evident that this step successfully separates clients 0, 1, and 8, 9. We then return to step 2 to continue the assessment until all communities/sub-communities meet the condition $Standard_{stop}$. 

\textbf{Step 4}: Clients are assigned to different communities or sub-communities through this process. Fig.~\ref{Fig11}e demonstrates that RL-CD accurately groups clients by deeply exploring data distribution relationships.

% Based on the results of the partitioning process, clients are assigned to different communities/sub-communities. While clients 0, 1, 8, and 9 belong to the same community, our algorithm precisely uncovers data distribution relationships and divides them into different sub-communities. Of course, if these sub-communities can be further divided, it would lead to even more specialized segmentation.

\subsubsection{Participant Selection}

After performing community detection, selecting clients from different communities/sub-communities can ensure data diversity and maximize $Div(S,t)$. Therefore, the goal of the participant selection is transformed into selecting clients to maximize $\sum_{i \in S} Util_{i}$ by comprehensively considering the stage-based memory, time, and data model.

\begin{comment}
    Algorithm \ref{client_selection} outlines the process of heterogeneity-aware client selection.

\end{comment}

\begin{comment}
    
\begin{algorithm}[h]  
  \caption{Stage-based heterogeneity-aware client selection.} 
  \label{client_selection}  
  \begin{algorithmic}[1]  
    \Require  

    $t$: Current stage;
    $Memory~Distribution$: Clients’ available memory distribution;
    $\Gamma$: Data size threshold;
    $\lambda$, $\varphi$: Prefer coefficient;
    Community set $C$: {$\{C_{1}, C_{2},..., C_{|C|}\}$};
    $\mathbf{t}_{t}^{i}$: Local training time at stage $t$;
    $I_{t,i}$: Local data loss;
    $\epsilon$: Exploration ratio.

    \Ensure  
      Client set $S$ with maximal $\Phi_{t}$ in current round at stage $t$. 

    \State $S^{*} \leftarrow$ empty set; 
    \State Current stage $t$;
    \State Calculate the $M_{train}(\Theta_{t})$ according to Eq.~\eqref{eq_memory2};
    \State Calculate the $Util$ of clients who satisfy Eq.~\eqref{Formulate18};

    \While{$|S^{*}| <(1-\epsilon)*|S|$}
        \For{each $C_{c} \in C$}
            \State append client $i \in C_{c}$ with maximal $Util_{i}$ to $S^{*}$;
            \State remove client $i$ from $C_{c}$;
        \If{$|S^{*}| \geq (1-\epsilon)*|S|$}
            \State \textbf{break}
        \EndIf
        \EndFor 
    \EndWhile
    \State Select $\epsilon *|S|$ clients from remaining clients who satisfy Eq.~\eqref{Formulate18} and add them to $S^{*}$;

    \State $S = S^{*}$;

    \State return client set $S$ ;
  \end{algorithmic}  
  
\end{algorithm}
\end{comment}

Client selection can be modeled as a multi-armed bandit problem~\cite{lai2021oort}, which maximizes rewards through exploration and exploitation. In the client selection problem, each client represents an arm of the bandit, with $Util$ as the reward, and the objective becomes to maximize the $Util$ of selected clients. SmartFreeze strategically exploits participants with a known high $Util$ at a ratio of $1-\epsilon$, while also exploring potential high $Util$ participants at a ratio of $\epsilon$, where $\epsilon \in [0,1]$. This is for the reason that not all clients participate in each training round, so the client's $Util$ may not be promptly updated. 

% Furthermore, the $Util$ of clients changes dynamically in different rounds of different training stages.

% The client selection can be modelled as a multi-armed bandit problem~\cite{lai2021oort}, where each client represents an arm of the bandit and the $Util$ serves as the reward.
% SmartFreeze strategically exploits participants with a known high $Util$ ratio of $1-\epsilon$ while exploring potential high $Util$ participants with a ratio of $\epsilon$, where $\epsilon \in [0,1]$. This approach accounts for not all clients participating in each round, so the client's $Util$ may not be promptly updated. Furthermore, the $Util$ of clients changes dynamically in different rounds of different training stages.

\begin{comment}
    
\section{Implementation}

We implement SmartFreeze in Python, utilizing the PyTorch package, resulting in a total of 3406 lines of code. In our simulation experiments, we harness the power of an A100 computing cluster to simulate a large-scale cross-device FL scenario. Simultaneously, we deploy local training process on embedded devices, running as a background service with PyTorch. We employ the Monsoon Power Monitor~\cite{Monsoon} to assess computation time and energy consumption, and used htop~\cite{htop} to monitor memory usage.
\end{comment}
\vspace{-5pt}
\section{Evaluation}

In this section, we first introduce the experimental methodology and baselines. Then, we discuss the corresponding experimental results. 
\vspace{-4pt}
\subsection{Experiment Setup}\label{sub_exp_set}

We evaluate SmartFreeze with both simulation and hardware testbeds. This hybrid testbed is used to cross-validate the performance and effectiveness of corresponding models and system design. Specifically, we establish an on-device FL system based on mobile devices with different hardware configurations, including Raspberry Pi 4B, NVIDIA Jetson Nano, NVIDIA Jetson TX2, and five types of mobile phones (Honor 8 Lite, Lenovo A3580, ZTE N928Dt, MI 4C and Nexus 6). The local training process in the mobile devices is implemented based on DL4J~\cite{dl4j} and running as a background service. Communication between the devices and the central server is established using the client-server architecture. We employ the Monsoon Power Monitor~\cite{Monsoon} to assess energy consumption and use htop~\cite{htop} to monitor memory usage.

We establish two federated training tasks: one involving 200 clients for the CIFAR10 dataset and the other with 100 clients for the CIFAR100 dataset. At the same time, we consider two memory distribution scenarios~\cite{memory_ram}: 1) When there are many applications running in the background (such as when running Google Chrome, which occupies 2.2GB of memory~\cite{memory_ram}), the available memory of the device decreases due to resource contention. We use this memory distribution to train ResNet10 and ResNet18 models. 2) When the resource contention is relatively low, the available memory of the device is high. We use this memory distribution to train the VGG11$\_$bn and VGG16$\_$bn models. For each training round, 20 clients are selected. The datasets are assigned to the clients with IID and Non-IID strategies. The Non-IID is set up based on Dirichlet distribution using a concentration parameter set to $\alpha=1$.

\begin{figure*}[!h]
    \centering

    \begin{subfigure}{0.23\textwidth}
        \centering
        \includegraphics[width=\linewidth]{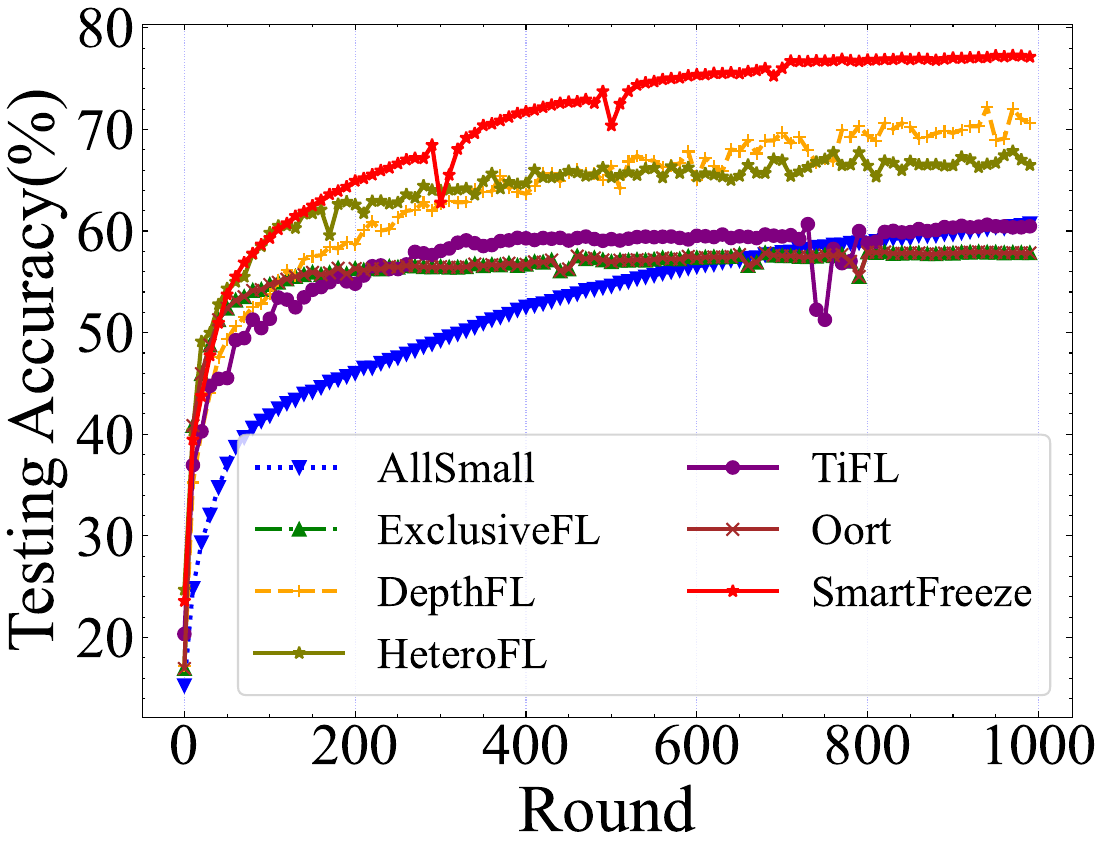}
        \caption{CIFAR10, IID.}
        \label{resulte}
    \end{subfigure}\hfill
    \begin{subfigure}{0.23\textwidth}
        \centering
        \includegraphics[width=\linewidth]{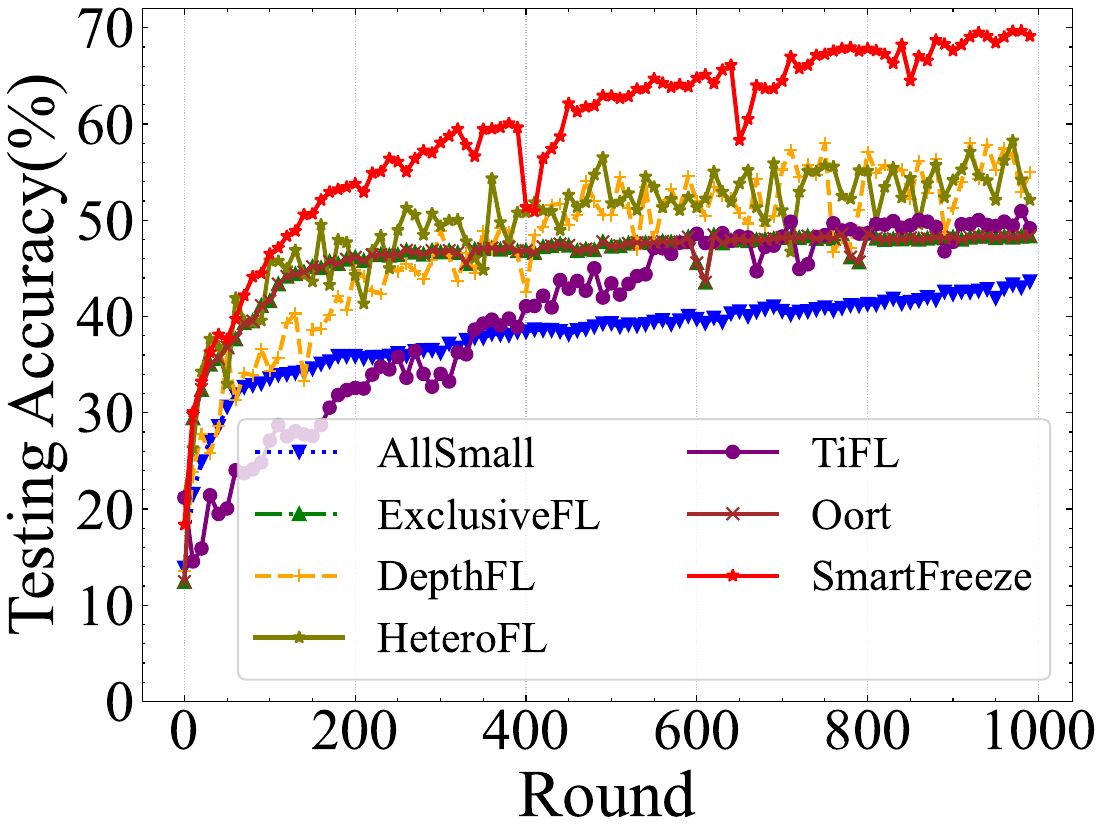}
        \caption{CIFAR10, Non-IID.}
        \label{resultf}
    \end{subfigure}\hfill
    \begin{subfigure}{0.23\textwidth}
        \centering
        \includegraphics[width=\linewidth]{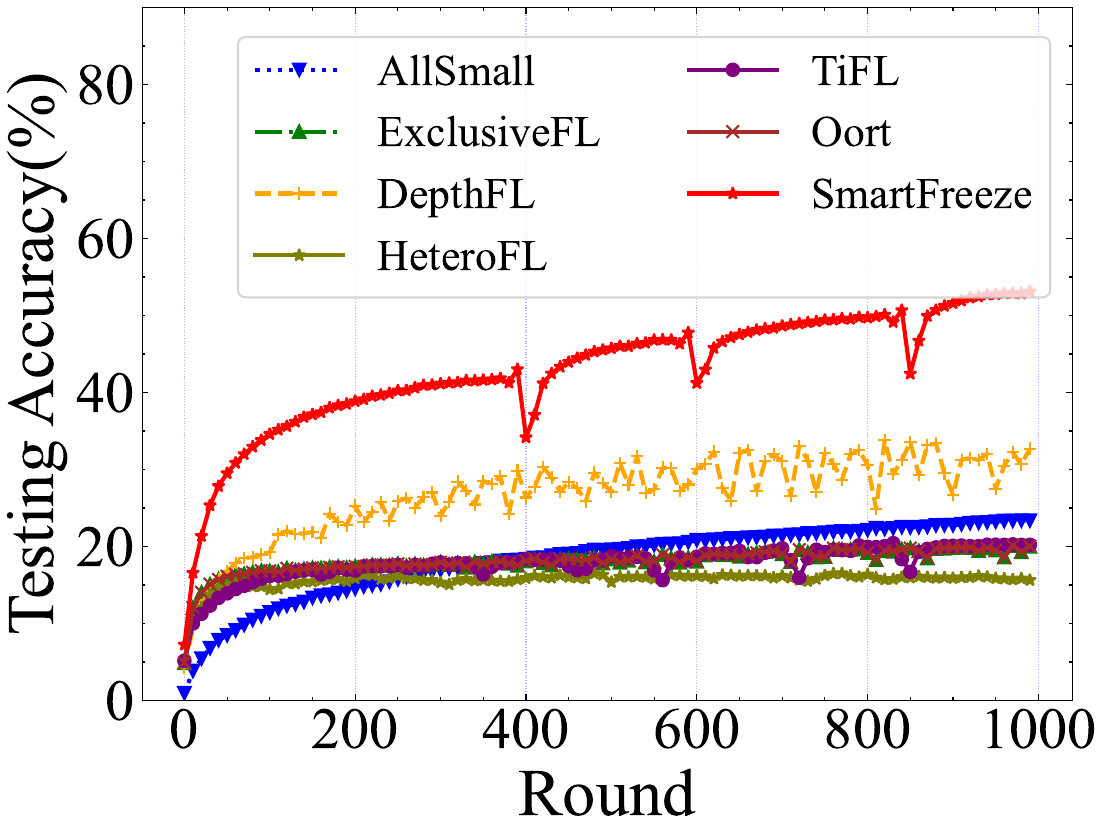}
        \caption{CIFAR100, IID.}
        \label{resultg}
    \end{subfigure}\hfill
    \begin{subfigure}{0.23\textwidth}
        \centering
        \includegraphics[width=\linewidth]{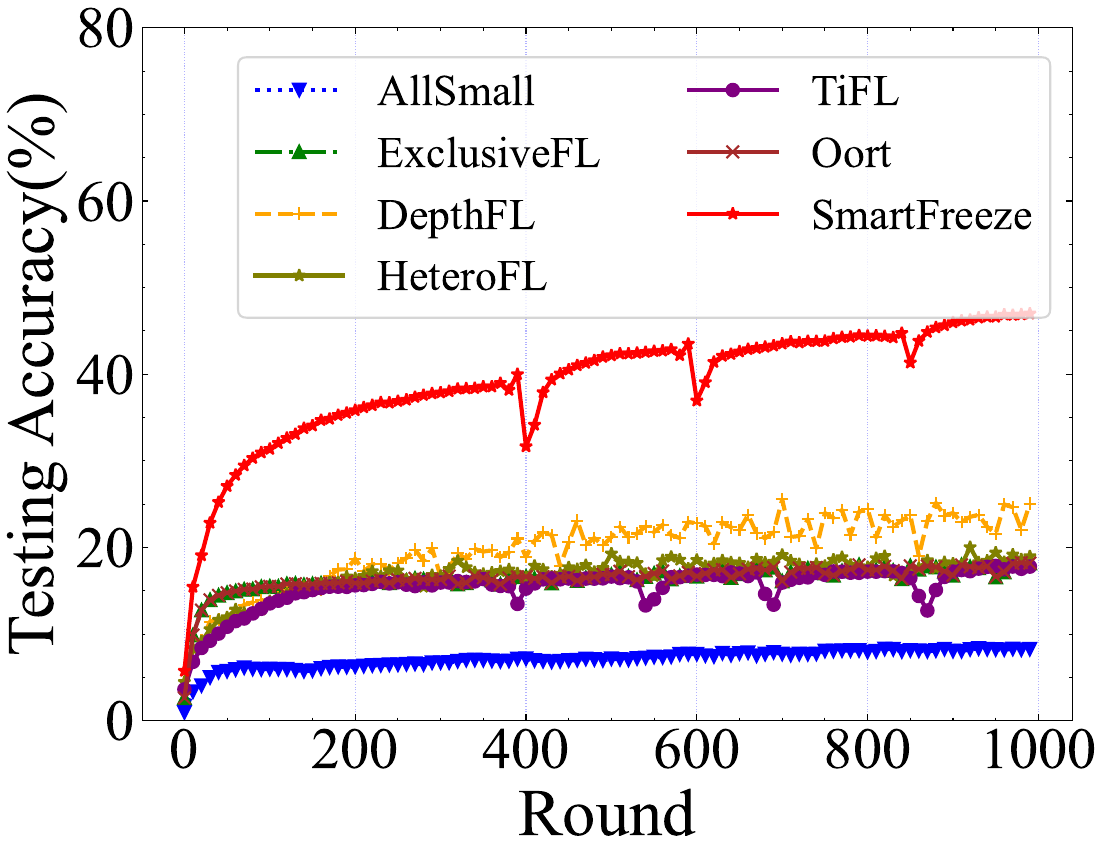}
        \caption{CIFAR100, Non-IID.}
        \label{resulth}
    \end{subfigure}
    \caption{Performance comparison of FL methods to train the ResNet10 model.}
    \label{Fig18_res10}
    \vspace{-12pt}
\end{figure*}

\begin{figure*}[!h]
    \centering
    \begin{subfigure}{0.23\textwidth}
        \centering
        \includegraphics[width=\linewidth]{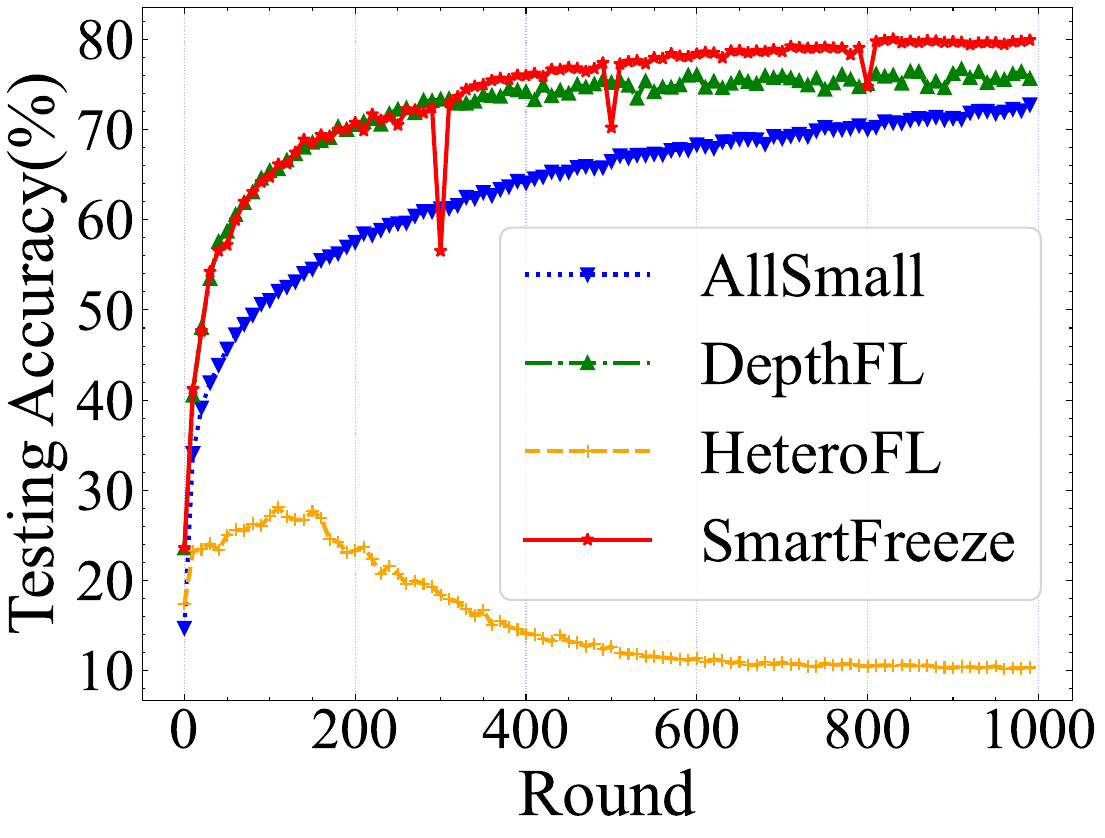}
        \caption{CIFAR10, IID.}
        \label{resulta}
    \end{subfigure}\hfill
    \begin{subfigure}{0.23\textwidth}
        \centering
        \includegraphics[width=\linewidth]{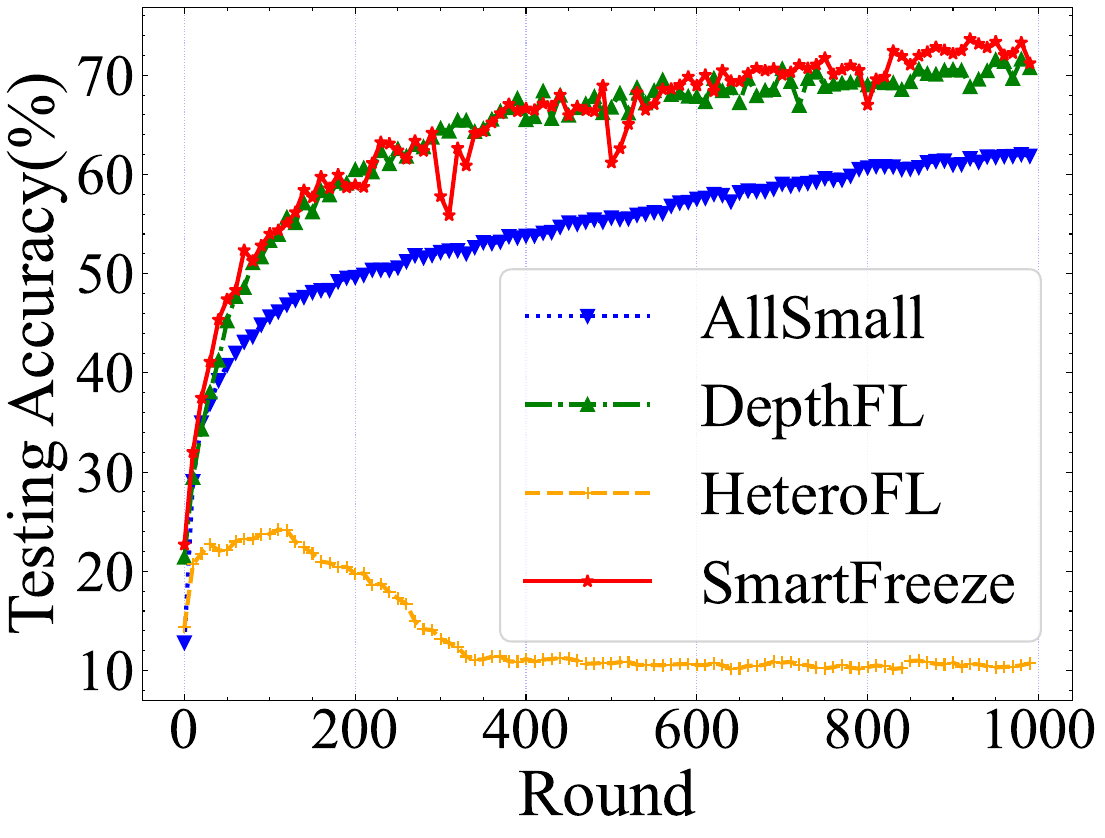}
        \caption{CIFAR10, Non-IID.}
        \label{resultb}
    \end{subfigure}\hfill
    \begin{subfigure}{0.23\textwidth}
        \centering
        \includegraphics[width=\linewidth]{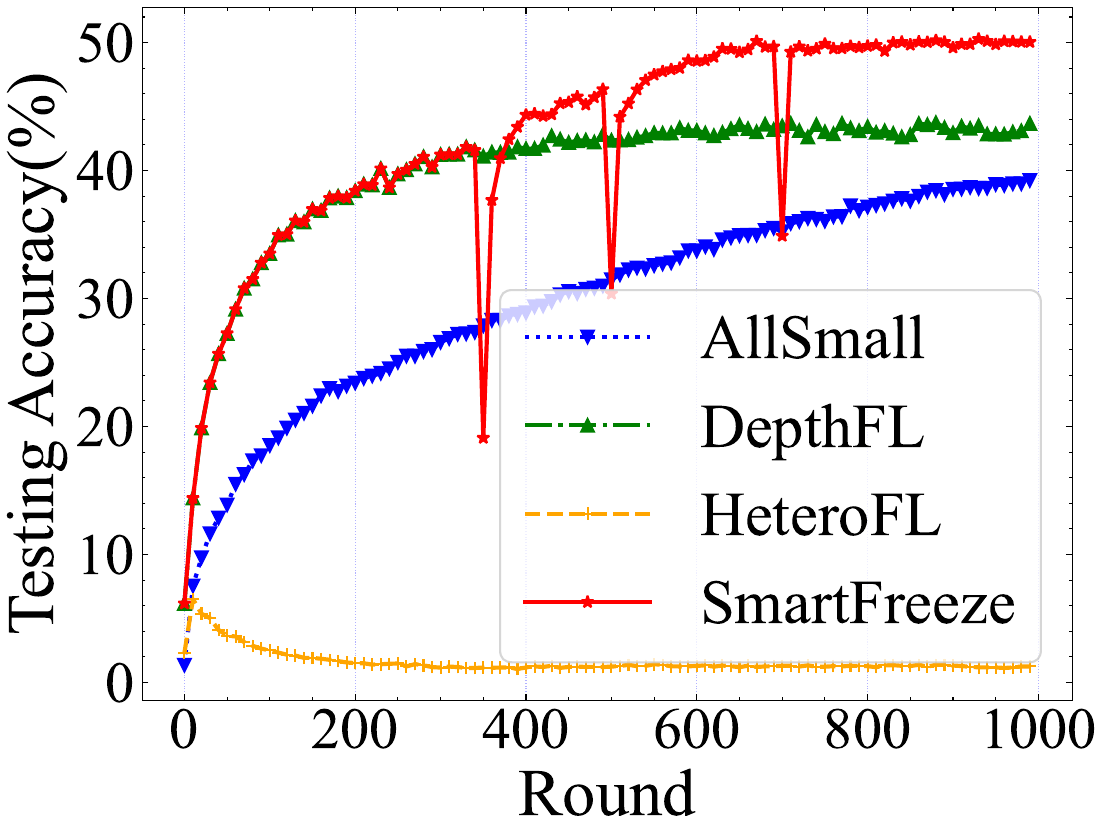}
        \caption{CIFAR100, IID.}
        \label{resultc}
    \end{subfigure}\hfill
    \begin{subfigure}{0.23\textwidth}
        \centering
        \includegraphics[width=\linewidth]{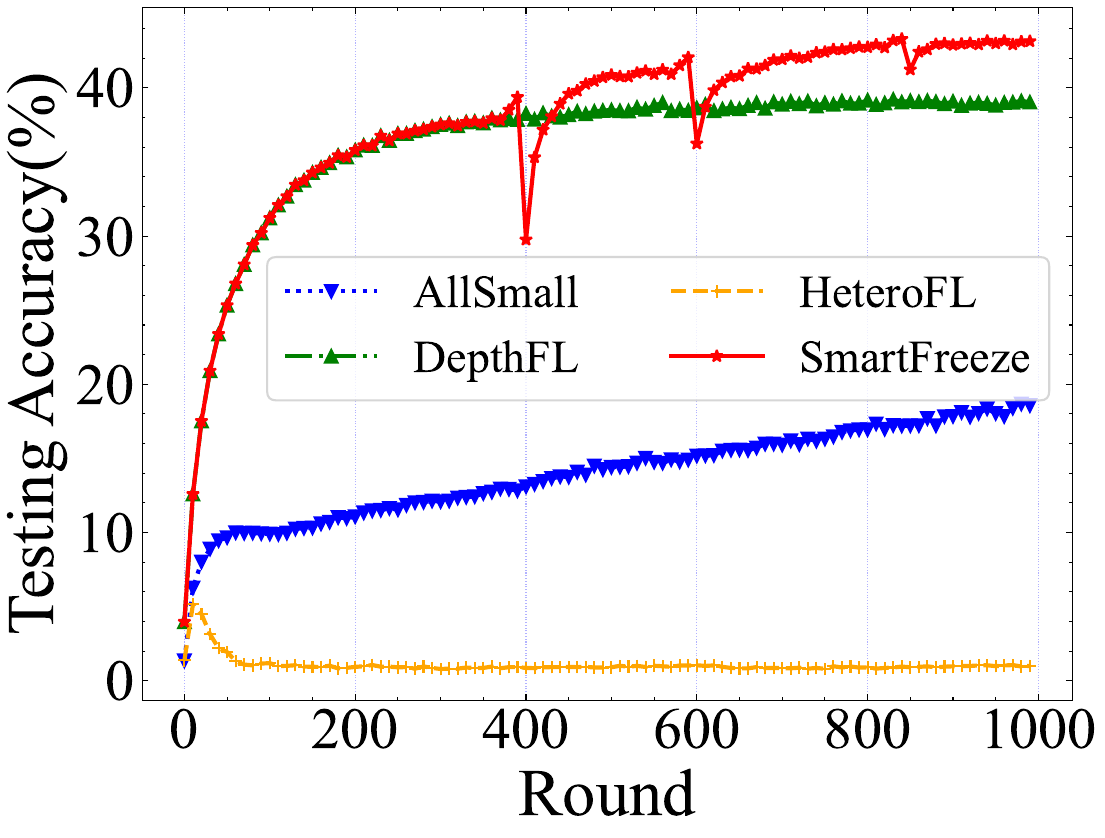}
        \caption{CIFAR100, Non-IID.}
        \label{resultd}
    \end{subfigure} \hfill
    
    \caption{Performance comparison of FL methods to train the ResNet18 model. Due to the inability of any device to afford the training of the full model, ExclusiveFL, TiFL, and Oort are inoperative in such situations.}
    \label{Fig18_res18}
    \vspace{-8pt}
\end{figure*}

\begin{table*}[!h]
  \footnotesize
  \centering
  \caption{Performance comparison of FL methods to train the VGG11$\_$bn and VGG16$\_$bn models. VGG11 and VGG16 represent the VGG11$\_$bn and VGG16$\_$bn, respectively. $PR$ represents the participation rate.}
  \label{performance_accvgg}
  \begin{tabular}{cc cccc cccc cc }
    \toprule
      &\multirow{2}{*}{\shortstack{Inclusive?}} &\multicolumn{2}{c}{\multirow{2}{*}{CIFAR10 (IID)}} &\multicolumn{2}{c}{\multirow{2}{*}{CIFAR10 (Non-IID)}} &\multicolumn{2}{c}{\multirow{2}{*}{CIFAR100 (IID)}} &\multicolumn{2}{c}{\multirow{2}{*}{CIFAR100 (Non-IID)}}&\multicolumn{2}{c}{\multirow{2}{*}{PR}} \\
      \\      &&VGG11 &VGG16&VGG11 &VGG16&VGG11 &VGG16&VGG11 &VGG16&VGG11 &VGG16  \\
    \midrule
    AllSmall &Yes& 82.6\%&79.2\% & 76.7\% &70.4\%& 50.2\%& 39.4\%& 42.3\% &30.8\%&100\%&100\%  \\
    \midrule
    ExclusiveFL&No& 84.5\%&NA & 82.5\% &NA&50.3\%&NA&48.9\% &NA&20\%&0\%\\
    \midrule

    DepthFL&No&86.3\%&77.0\% & 83.4\% &69.3\%&56.7\%&40.2\%&53.4\% &36.1\%&41\%&34\% \\
    \midrule
    HeteroFL&Yes&80.9\%&10.5\% & 75.2\% &10.2\%& 50.8\%&1.4\%& 43.6\% &1.1\%&100\%&100\% \\
    \midrule 
    TiFL&No&84.4\%&NA &73.8\% &NA& 51.3\%&NA& 49.3\% &NA&20\%&0\% \\
    \midrule 
    Oort&No&85.1\%&NA & 83.3\% &NA&53.2\%&NA& 46.2\% &NA&20\%&0\% \\
    \midrule 
    
    SmartFreeze&Yes&\textbf{88.4\%}&\textbf{83.1\%} & \textbf{84.9\%} &\textbf{74.8\%}& \textbf{63.1\%}&\textbf{48.7\%}& \textbf{59.7\%} &\textbf{44.2\%}&100\%&100\%\\
    \bottomrule
  \end{tabular}
  \vspace{-15pt}
\end{table*}

\textbf{Baselines.}
The following baselines are adopted for evaluation purposes: 
\begin{itemize}
    \item \textit{AllSmall~\cite{liu2022no}} is a naive baseline that scales the channels of the convolutional layer based on the minimum client memory to accommodate clients with different memory capacities for training.
    
    % \item AllHigh: A naive baseline where all clients have the maximum client's available memory (in this case 80\% of the memory required to train the entire model).

    % \item Naive layer freezing: A naive baseline, layers are frozen from front to back based on the client's available memory.

    \item \textit{ExclusiveFL~\cite{liu2022no}} only allows clients with sufficient memory capacity to train the full model to participate in training, excluding clients with insufficient memory.

    \item \textit{DepthFL}~\cite{kim2022depthfl} segments the model into sub-models of varying depths, distributing them to clients according to their memory capacity.
    % In the case of memory constraints, ExclusiveFL and DepthFL degenerate into a single algorithm.

     % \item \textit{HeteroFL}~\cite{diao2020heterofl} scales the channels of convolutional layers to adapt to different memory capacities, thereby facilitating training across heterogeneous devices.

    \item \textit{HeteroFL}~\cite{diao2020heterofl} facilitates training across heterogeneous devices by  scaling the channels of convolutional layers to cater to diverse memory constraints.
    % targeting optimal adaptation to varied memory limitations.
    
    % \item Federated Dropout\cite{caldas2018expanding}: In contrast to HeteroFL, the selection of channels for each clients in every round changes.
    
    \item  \textit{TiFL}~\cite{chai2020tifl} stratifies devices according to training time, selecting devices from specific tiers each round and adjusting the selection probability for subsequent rounds based on accuracy.

    \item \textit{Oort}~\cite{lai2021oort} aims to enhance time-to-accuracy by considering both data and system heterogeneity during client selection.
    
\end{itemize}

\vspace{-7pt}
\subsection{End-to-End Evaluation}
\subsubsection{Effectiveness}
We present benchmark evaluation results of SmartFreeze in end-to-end training models. Fig.~\ref{Fig18_res10} and Fig.~\ref{Fig18_res18} illustrate the accuracy convergence curves of ResNet10 and ResNet18 under various algorithms, respectively. The X-axis represents the training rounds, while the Y-axis represents the testing accuracy. Table~\ref{performance_accvgg} presents accuracy results for VGG series models.

Fig.~\ref{resulte} illustrates that SmartFreeze significantly outshines baselines on the CIFAR10 dataset. When compared to AllSmall, SmartFreeze delivers a 16.8\% performance improvement. The primary limitation of AllSmall stems from the model size being determined by the client with the smallest memory capacity, consequently impairing feature extraction capability. SmartFreeze enhances feature extraction capability through progressive model growth. In contrast to ExclusiveFL, SmartFreeze exhibits a performance improvement of 19.6\%. This can be attributed to ExclusiveFL's stringent inclusion criteria, allowing only high-memory clients to participate, resulting in a participation rate of only 3\%. DepthFL employs self-distillation to facilitate information transfer among blocks. However, its limited 28\% participation rate results in a 5\% decrease in performance compared to SmartFreeze. Due to the disruption of the model architecture, HeteroFL results in a performance degradation of 9.2\%. Both TiFL and Oort, having a mere 3\% participation rates, endure an accuracy reduction of 16.9\% and 19.6\%, respectively. Fig.~\ref{resultf} highlights a similar trend on CIFAR10 (Non-IID), with SmartFreeze achieving accuracy improvements of 26.2\%, 22.3\%, 7.6\%, 11\%, 19.5\%, and 22.3\% compared to baselines. When examining the CIFAR100 dataset in Fig.\ref{resultg} and Fig.\ref{resulth}, SmartFreeze consistently excels in both IID and Non-IID environments, with performance improvements of 18.7\%-36.4\% and 19.3\%-38.5\%, respectively. Compared with the results obtained on the CIFAR10 and CIFAR100, we can see that SmartFreeze is more effective in handling complex tasks.

% \begin{figure*}[!h]
%     \centering
%     \begin{subfigure}{0.23\textwidth}
%         \centering
%         \includegraphics[width=\linewidth]{evaluation/IID_cifar10_time.pdf}
%         \caption{ToA (CIFAR10).}
%         \label{time1}
%     \end{subfigure}\hfill
%     \begin{subfigure}{0.23\textwidth}
%         \centering
%         \includegraphics[width=\linewidth]{evaluation/IID_cifar100_time.pdf}
%         \caption{ToA (CIFAR100).}
%         \label{time2}
%     \end{subfigure}\hfill
%     \begin{subfigure}{0.23\textwidth}
%         \centering
%         \includegraphics[width=\linewidth]{evaluation/IID_cifar10_energy.pdf}
%         \caption{EoA (CIFAR10).}
%         \label{energy1}
%     \end{subfigure}\hfill
%     \begin{subfigure}{0.23\textwidth}
%         \centering
%         \includegraphics[width=\linewidth]{evaluation/IID_cifar100_energy.pdf}
%         \caption{EoA (CIFAR100).}
%         \label{energy2}
%     \end{subfigure} \hfill
%     \caption{Efficiency comparison of various schemes to train the ResNet18 model (CIFAR10/100, IID). ToA: time to accuracy, EoA: energy to accuracy.}
%     \label{efficiency}
%     \vspace{-18pt}
    
% \end{figure*}

Fig.~\ref{Fig18_res18} depicts the results of training the ResNet18 model. Due to the increased model complexity, the memory wall disables the ExclusiveFL, TiFL, and Oort frameworks, with no client possessing enough memory to train the full model. Conversely, SmartFreeze consistently demonstrates excellent performance, achieving accuracy improvements ranging from 1.5\% to 80.2\% on both datasets. This can be attributed to the progressive training paradigm, which, on the one hand, adapts to increased model complexity and, on the other hand, involves more low-memory devices. For VGG series models, from Table~\ref{performance_accvgg}, it can be observed that VGG11$\_$bn exhibits a similar trend to ResNet10, with an improvement in accuracy ranging from 1.5\% to 17.4\% compared to other algorithms. VGG16$\_$bn shows performance similar to ResNet18, with an accuracy improvement ranging from 3.9\% to 83.1\%.

\subsubsection{Efficiency}
By intelligently selecting clients, SmartFreeze enhances 
model effectiveness and system efficiency at the same time. Through extensive experiments, SmartFreeze significantly outperforms in terms of time-to-accuracy and energy-to-accuracy metrics on the CIFAR10 and CIFAR100 datasets. This can be attributed to the stage-based progressive training paradigm, which involves training only a subset of parameters at each stage.  
Consequently, both forward and backward propagation computation demands are substantially diminished compared to the full model training, resulting in reduced time and energy costs per round and yielding a speedup of up to 2.02$\times$. It is pertinent to mention that the overhead introduced by the concatenated output module for each block is insubstantial, with a mean memory overhead of 2.8\% and a computation overhead of 7.3\%.

\vspace{-6pt}
\subsection{Performance Breakdown}

\addtolength{\topmargin}{0.06in}

\subsubsection{\textbf{Effectiveness of RL-CD Algorithm}}

% RL-CD algorithm is primarily used to uncover the data distribution relationships among clients, ensuring data diversity in client selection. 

\begin{figure}[h]
    \vspace{-5pt}
    \centering
    \begin{subfigure}{0.23\textwidth}
        \centering
        \includegraphics[width=\linewidth]{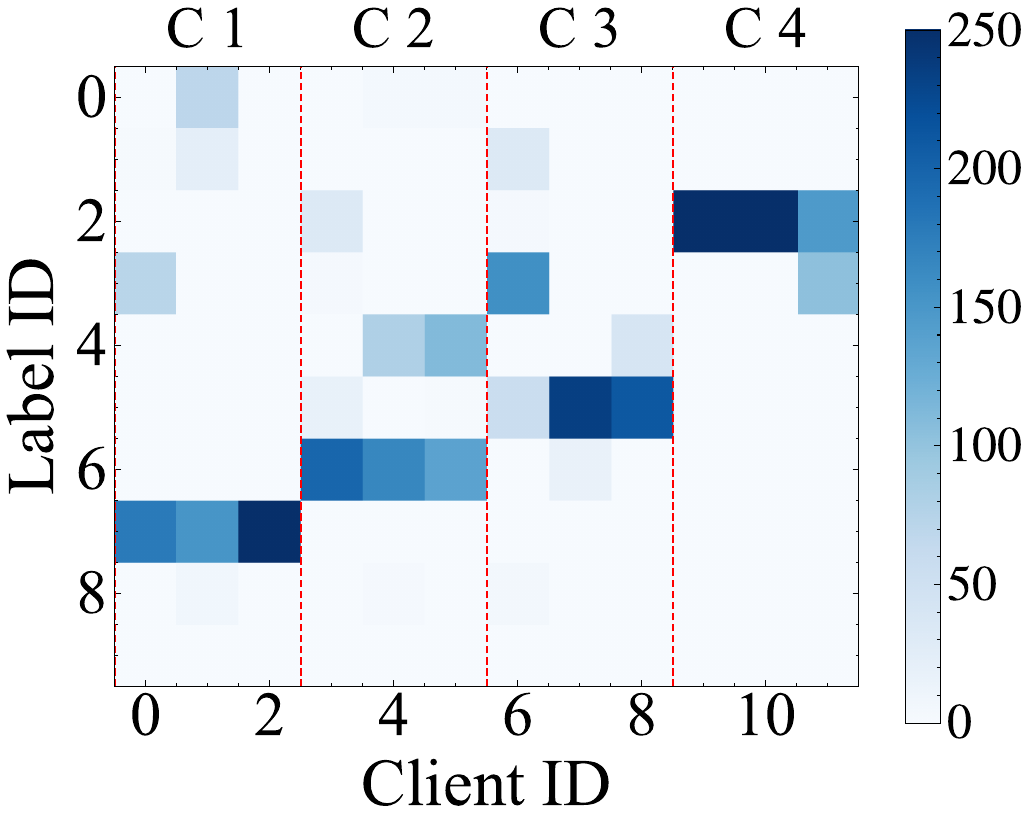}
        \caption{Community results.}
        \label{community1}
    \end{subfigure}\hfill
    \begin{subfigure}{0.23\textwidth}
        \centering
        \includegraphics[width=\linewidth]{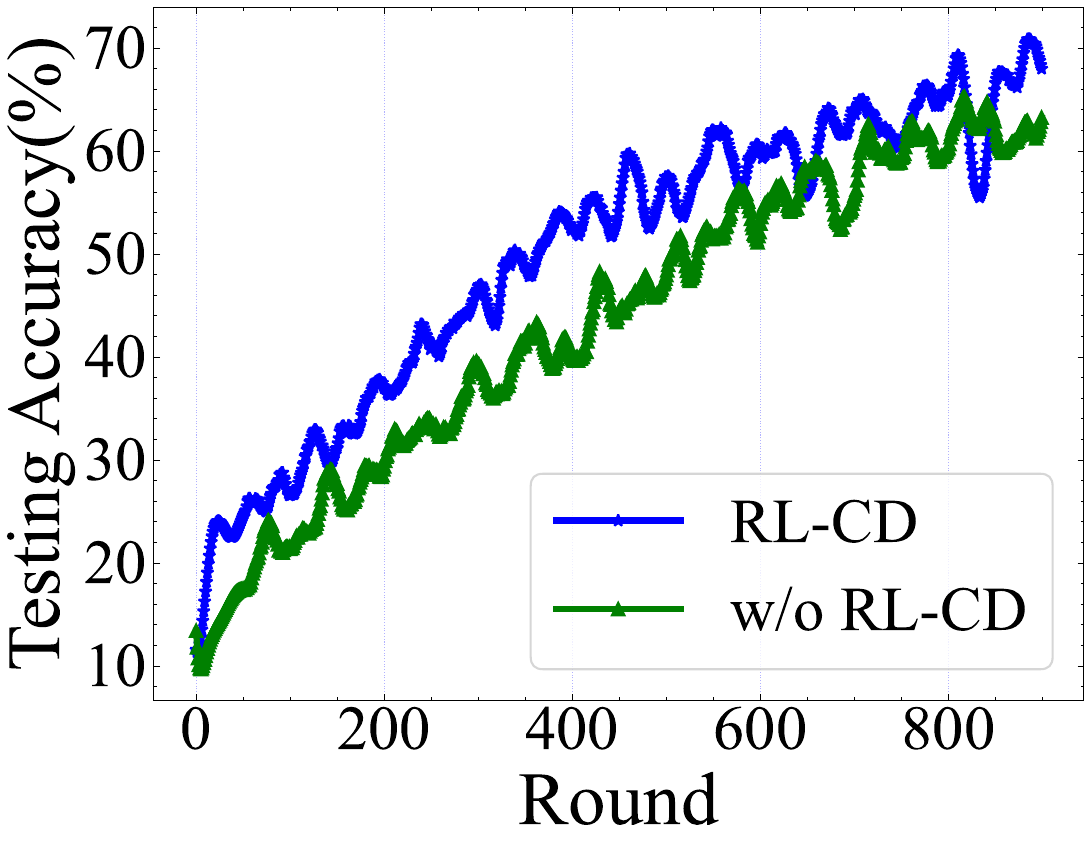}            
        \caption{Performance comparison.}
        \label{community2}
    \end{subfigure}
    \caption{Performance breakdown of RL-CD algorithm. C1-C4 represent four communities. w/o RL-CD means without RL-CD algorithm.}
    \label{community}
    \vspace{-18pt}
\end{figure}

To underscore the importance of RL-CD, we conduct a more pronounced Non-IID partitioning on CIFAR10. Fig.~\ref{community} displays the detected communities using the RL-CD algorithm and the corresponding convergence results. We select four communities (C1-C4) from the partitioned sets, showcasing the data distribution of three clients within each community. Fig.~\ref{community1} shows that clients within the same community share similar data distributions, while significant differences exist between different communities. The effectiveness of RL-CD in accelerating model convergence is demonstrated in Fig.~\ref{community2}, resulting in a 7.4\% accuracy improvement. 

% Therefore, our RL-CD algorithm can accurately capture client data distribution relationships, thereby improving the final model performance.

% \begin{figure*}[h]
%     \centering
%     \begin{subfigure}{0.23\textwidth}
%         \centering
%         \includegraphics[width=\linewidth]{evaluation/IID_cifar10_perturbation.pdf}
%         \caption{CIFAR10, IID.}
%         \label{LP1}
%     \end{subfigure}\hfill
%     \begin{subfigure}{0.23\textwidth}
%         \centering
%         \includegraphics[width=\linewidth]{evaluation/non_IID_cifar10_perturbation.pdf}
%         \caption{CIFAR10, Non-IID.}
%         \label{LP2}
%     \end{subfigure}\hfill
%     \begin{subfigure}{0.23\textwidth}
%         \centering
%         \includegraphics[width=\linewidth]{evaluation/IID_cifar100_perturbation.pdf}
%         \caption{CIFAR100, IID.}
%         \label{LP3}
%     \end{subfigure}\hfill
%     \begin{subfigure}{0.23\textwidth}
%         \centering
%         \includegraphics[width=\linewidth]{evaluation/nonIID_cifar100_perturbation.pdf}
%         \caption{CIFAR100, Non-IID.}
%         \label{LP4}
%     \end{subfigure} \hfill
%     \caption{Block perturbation as a robust indicator reflecting block convergence status (ResNet18). Where \textit{stage} represents the block perturbation of the sub-model at the current stage. $Acc.$ represents the testing accuracy of the corresponding round. }
%     \label{LP} 
%     \vspace{-14pt}
% \end{figure*}

\subsubsection{\textbf{Effectiveness of Block Perturbation}} 

% Block perturbation is employed to assess the convergence status of each block at various stages, allowing them to be safely frozen, thereby triggering model growth to commence the next training stage. 

We employ the following settings for comparison: (a) only train the first block, (b) given that the last block is responsible for extracting more complex features, we freeze the earlier blocks prematurely and allocate more training rounds to the final block, (c) a naive method is to allocate training rounds based on the parameter quantity of different blocks. The total number of training rounds remains consistent across all experiments. 
Table~\ref{abla_free} indicates that improper layer freezing can adversely affect the model's performance. Compared to the other three settings, utilizing the block perturbation leads to accuracy improvements of 3.7\%, 4.1\%, and 7.5\%, respectively. Comparing the results of experiment (a) with those of (b) and (c) indicates that model growth through erroneous freezing leads to poorer performance compared to not applying model growth.

\begin{table}[!ht]
  \vspace{-3pt}
  \footnotesize
  \centering
  \caption{Performance comparison with and without block perturbation in FL on CIFAR10 (IID).}
  \label{abla_free}
  \begin{tabular}{ccccc}
    \toprule
    Experiment & With BP & a & b & c \\
    \midrule
    Accuracy & 80.1\% &76.4\% & 76.0\% & 72.6\% \\
  \bottomrule
\end{tabular}
\vspace{-15pt}
\end{table}

\subsection{{Hardware Evaluation}}

\begin{figure}[t]
    \vspace{-5pt}
    \centering  
    \begin{subfigure}{0.23\textwidth}
        \centering
        \includegraphics[width=\linewidth]{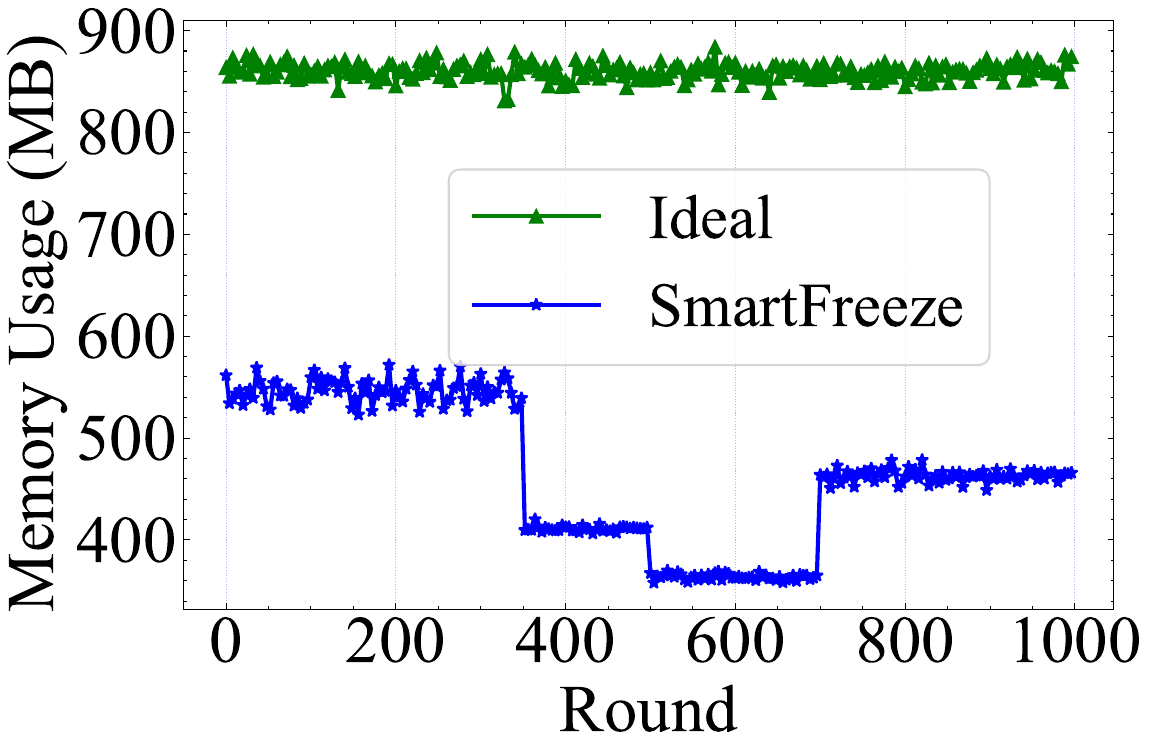}
        \caption{Raspberry Pi 4B.}
        \label{Fig19c}
    \end{subfigure}\hfill
        \begin{subfigure}{0.23\textwidth}
        \centering
        \includegraphics[width=\linewidth]{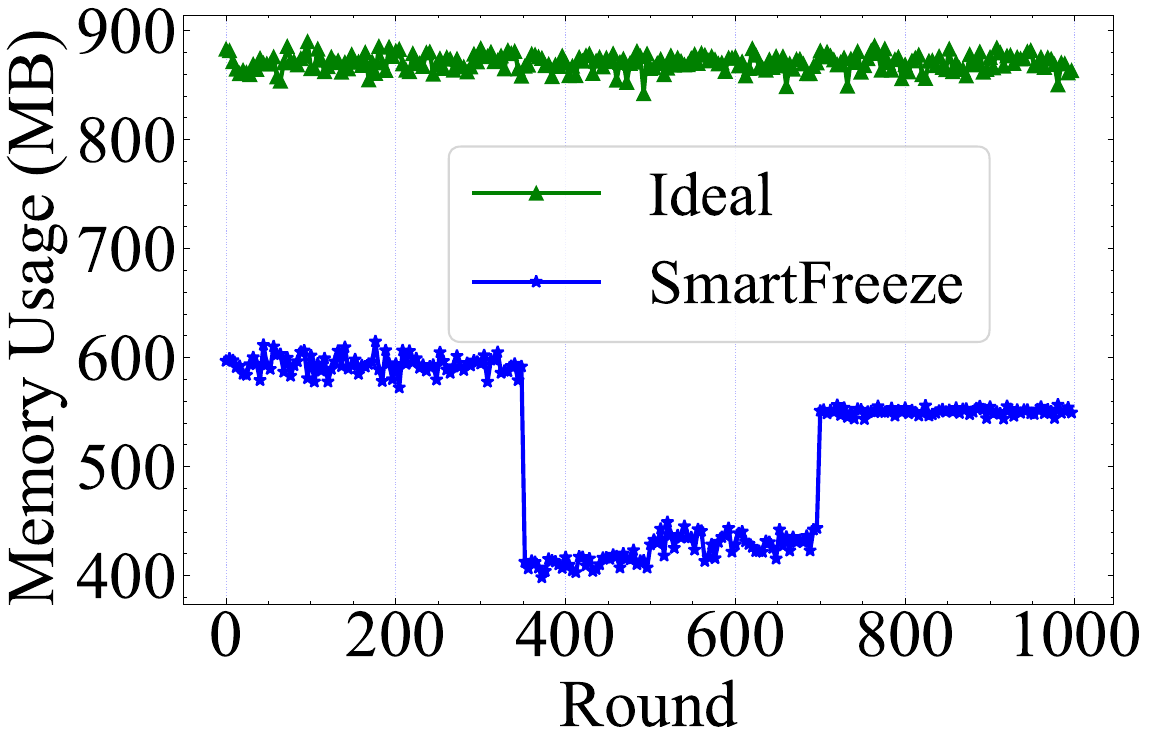}
        \caption{NVIDIA Jetson TX2.}
        \label{Fig19a}
    \end{subfigure}\hfill
    \caption{Hardware Memory Usage.}
    \label{Fig19}
    \vspace{-17pt}
\end{figure}

In our simulation testbed, we can achieve an average memory reduction of up to 82\%. 
To validate the memory efficiency of SmartFreeze on real devices, we select Raspberry Pi 4B and NVIDIA Jetson TX2 as experimental platforms, each equipped with 500 CIFAR100 data samples, to train the ResNet18 model. We use htop~\cite{htop} to monitor the peak memory usage. 
Fig.~\ref{Fig19} illustrates the memory usage of each device during training. The X-axis represents the training rounds, and the Y-axis represents the corresponding memory usage.

Fig.~\ref{Fig19} indicates that our method significantly reduces the memory requirements for training at each stage compared to the ideal scenario of training the full model on each device. When excluding the influence of the process itself, the memory requirements in the first stage can be reduced by 39.7\%-43.9\%, in the second stage by 63.2\%-64.8\%, in the third stage by 62.9\%-69.9\%, and in the fourth stage by 45.6\%-55.9\%. Therefore, SmartFreeze effectively decreases memory usage throughout the training process.

% \vspace{-6pt}
% \subsection{Discussion}

% \subsubsection{Output Module} This method of constructing output modules can retain more useful information during the training process. Through experiments, we find that compared to using fully connected layers directly as output modules, this method can improve accuracy by up to 4\%.

% \subsubsection{Communication Cost} Compared to end-to-end training of a full model, SmartFreeze only trains partial model parameters at each stage. In each training round, only the updated parameters need to be communicated, thus effectively reducing communication overhead. Compared to the ideal situation, SmartFreeze can effectively reduce communication overhead by up to 58\%.

% Compared to end-to-end training of the full model, SmartFreeze trains only a portion of the model parameters at each stage. In each round, only the updated parameters need to be communicated, thus effectively reducing communication overhead compared to training the full model.

% \subsubsection{Model Universality} To demonstrate the universality of SmartFreeze, we test it on the CIFAR100 dataset using the Vision Transformer (ViT). We divide the training of ViT into three stages, with each stage training four transformer encoders. Under the IID setting, it can achieve a classification accuracy of 53\%. Additionally, SmartFreeze can also adapt well to NLP-related models, as well as other optimizers.

% Although we only test ResNet and VGG series models in this paper, SmartFreeze can also adapt well to Vision Transformer (ViT)~\cite{dosovitskiy2020image} and NLP-related models~\cite{vaswani2017attention}~\cite{devlin2018bert}~\cite{radford2018improving}. For example, Deep Incubation~\cite{ni2023deep} accelerates the convergence of the ViT model by dividing it into different blocks for training. Additionally, SmartFreeze reduces memory overhead by releasing intermediate outputs and gradients of frozen layers, while also decreasing memory consumption by training only a portion of the model. Consequently, SmartFreeze can effectively adapt to different optimizers, such as Adam~\cite{kingma2014adam} and Adagrad~\cite{lydia2019adagrad}.

\vspace{-3pt}

\section{Related Work}
\noindent
\textbf{On-Device Training.} On-device training not only protects the privacy of local data but also adapts to the continuously evolving data. Study~\cite{huang2023elastictrainer} has shown that deploying pre-trained models directly without local updates can result in an inference accuracy decrease by up to 20\%. Therefore, on-device training becomes crucial. ElasticTrainer~\cite{huang2023elastictrainer} improves computation efficiency by transitioning from a layer-based structure to a tensor-based structure, enabling prioritized training of critical tensors within time constraints. Additionally, Miro~\cite{ma2023cost} optimizes cost-effectiveness based on resource availability for continuous learning.

% The rise of on-device training is driven by advancements in chip technology, offering data privacy and adaptability to evolving data streams. Research has indicated that deploying pre-trained models on devices without updates can reduce inference accuracy by up to 20\%. Therefore, on-device training becomes crucial. 
% The ElasticTrainer~\cite{huang2023elastictrainer} enhances computational efficiency by transitioning from a layer-based structure to a tensor-based structure, enabling the prioritized training of crucial tensors within time constraints. Additionally, Miro~\cite{ma2023cost} conducts continuous learning while optimizing cost-effectiveness based on available resource allocation.

\noindent
\textbf{Memory-Constrained FL.} The implementation of FL faces challenges related to memory heterogeneity, especially in cross-device FL~\cite{wu2024breaking}\cite{fu2022federated}~\cite{li2022one}. Many algorithms assume uniform client model size~\cite{wang2023fedins2}\cite{tian2022harmony}, but this assumption often does not hold in practice. Training large models typically only involves high-memory devices, neglecting the valuable data from low-memory devices. HeteroFL~\cite{diao2020heterofl} divides the model into different complexities by scaling the channels of convolutional layers, but this compromises the model architecture. FedMD~\cite{li2019fedmd} tailors the model complexity to each client but requires additional public datasets, which is not feasible in FL.

% In contrast, SmartFreeze provides a flexible method for training large models without imposing additional burdens.

\vspace{-3pt}
\section{Conclusion}

% In this paper, we have introduced the progressive layer freezing FL framework based on stage training, which addresses the challenge of memory heterogeneity in real-world FL scenarios. Our approach facilitates the training of a large global model while effectively harnessing data information from low-memory devices. Consequently, we mitigate the performance degradation issues associated with the use of smaller models or the exclusion of low-memory devices. Furthermore, we have proposed a novel metric for assessing layer convergence. 

% Extensive experiments were conducted in both simulation and real-device settings. SmartFreeze achieves significant performance improvements, ranging from 1.9\% to an impressive 80.1\%, when compared to other algorithms. Importantly, these enhancements do not impact training time or result in increased energy consumption. Furthermore, in terms of memory usage, SmartFreeze excels, reducing it by an impressive 79.4\% to 81.9\% compared to ideal conditions. In comparison to other algorithms (excluding AllSmall), it still manages a notable reduction in memory usage, ranging from 6.6\% to 13.8\%. Even in real-world testing on embedded devices, SmartFreeze maintains its efficiency by reducing memory usage by an impressive 42.8\% to 52.1\% compared to the ideal conditions. Our findings underscore the practical significance of SmartFreeze in addressing memory heterogeneity challenges in FL, ensuring efficient utilization of resources and data while maintaining data privacy and model performance.

This paper introduces SmartFreeze, a framework designed to effectively reduce peak memory footprint while accounting for system and statistical heterogeneity. SmartFreeze employs a progressive training paradigm and proposes a block perturbation metric to evaluate the training progress of blocks to securely freeze them as they converge. This significantly saves computation resources for frozen blocks and the corresponding memory space used for storing intermediate outputs and gradients. Additionally, to address the statistical and system heterogeneity, we propose a graph-based community detection algorithm to choose the \emph{right} devices to participate in the training for each block. We conduct extensive experiments to evaluate the effectiveness of SmartFreeze on both simulation and hardware testbeds. The results demonstrate that SmartFreeze effectively reduces average memory usage by up to 82\%. Moreover, it improves the model accuracy by up to 83.1\% and accelerates the training process up to 2.02$\times$.

% In this paper, we introduce SmartFreeze, a framework designed to effectively reduce peak memory demands while accounting for system and statistical heterogeneity. SmartFreeze employs a progressive training paradigm, securely freezing blocks as they converge and subsequently triggering the training of the next block. This significantly saves computational resources for frozen blocks and the corresponding memory space used for storing intermediate outputs and gradients. Additionally, to address the statistical and system heterogeneity, we propose a graph-based client selection algorithm to choose the \emph{right} devices for each block during training. We conduct extensive experiments to evaluate the effectiveness of SmartFreeze on simulation and hardware testing platforms. The results demonstrate that, SmartFreeze effectively reduces memory usage up to 82\%. Moreover, it improves the model performance up to 80.2\% and accelerates the training process up to 2.02 $\times$. 

\section*{Acknowledgment}
This paper is supported the Science and Technology Development Fund of Macau SAR (File no. 0081/2022/A2, 0123/2022/AFJ), MYRG-GRG2023-00211-10TSC-UMDF, SRG2022-00010-10TSC. Please ask Dr. ChengZhong Xu (czxu@um.edu.mo) and  Dr. Li Li (llili@um.edu.mo) for correspondence.
\vspace{-2pt}
\bibliography{ref2}

\begin{thebibliography}{10}
\providecommand{\url}[1]{#1}
\csname url@samestyle\endcsname
\providecommand{\newblock}{\relax}
\providecommand{\bibinfo}[2]{#2}
\providecommand{\BIBentrySTDinterwordspacing}{\spaceskip=0pt\relax}
\providecommand{\BIBentryALTinterwordstretchfactor}{4}
\providecommand{\BIBentryALTinterwordspacing}{\spaceskip=\fontdimen2\font plus
\BIBentryALTinterwordstretchfactor\fontdimen3\font minus \fontdimen4\font\relax}
\providecommand{\BIBforeignlanguage}[2]{{%
\expandafter\ifx\csname l@#1\endcsname\relax
\typeout{** WARNING: IEEEtran.bst: No hyphenation pattern has been}%
\typeout{** loaded for the language `#1'. Using the pattern for}%
\typeout{** the default language instead.}%
\else
\language=\csname l@#1\endcsname
\fi
#2}}
\providecommand{\BIBdecl}{\relax}
\BIBdecl

\bibitem{mcmahan2017communication}
{McMahan, Brendan} \emph{et~al.}, ``Communication-efficient learning of deep networks from decentralized data.''\hskip 1em plus 0.5em minus 0.4em\relax PMLR, 2017, pp. 1273--1282.

\bibitem{he2016deep}
{He, Kaiming} \emph{et~al.}, ``Deep residual learning for image recognition,'' in \emph{CVPR}, 2016, pp. 770--778.

\bibitem{memory_ram}
\BIBentryALTinterwordspacing
``How much ram does your android phone really need in 2023?'' 2023. [Online]. Available: \url{https://www.androidauthority.com}
\BIBentrySTDinterwordspacing

\bibitem{wang2022melon}
{Wang, Qipeng} \emph{et~al.}, ``Melon: Breaking the memory wall for resource-efficient on-device machine learning,'' in \emph{MobiSys}, 2022, pp. 450--463.

\bibitem{gim2022memory}
{Gim, In} \emph{et~al.}, ``Memory-efficient dnn training on mobile devices,'' in \emph{MobiSys}, 2022, pp. 464--476.

\bibitem{cao2022framework}
{Cao, Qinglei} \emph{et~al.}, ``A framework to exploit data sparsity in tile low-rank cholesky factorization,'' in \emph{IPDPS}.\hskip 1em plus 0.5em minus 0.4em\relax IEEE, 2022, pp. 414--424.

\bibitem{huang2020swapadvisor}
{Huang, Chien-Chin} \emph{et~al.}, ``Swapadvisor: Pushing deep learning beyond the gpu memory limit via smart swapping,'' in \emph{ASPLOS}, 2020, pp. 1341--1355.

\bibitem{li2019fedmd}
{Li, Daliang} \emph{et~al.}, ``Fedmd: Heterogenous federated learning via model distillation,'' \emph{arXiv preprint arXiv:1910.03581}, 2019.

\bibitem{zhang2022fedzkt}
{Zhang, Lan} \emph{et~al.}, ``Fedzkt: Zero-shot knowledge transfer towards resource-constrained federated learning with heterogeneous on-device models,'' in \emph{ICDCS}.\hskip 1em plus 0.5em minus 0.4em\relax IEEE, 2022, pp. 928--938.

\bibitem{diao2020heterofl}
{Diao, Enmao} \emph{et~al.}, ``Heterofl: Computation and communication efficient federated learning for heterogeneous clients,'' \emph{arXiv preprint arXiv:2010.01264}, 2020.

\bibitem{yang2022partial}
{Yang, Tien-Ju} \emph{et~al.}, ``Partial variable training for efficient on-device federated learning,'' in \emph{ICASSP}.\hskip 1em plus 0.5em minus 0.4em\relax IEEE, 2022, pp. 4348--4352.

\bibitem{hinton2015distilling}
{Hinton, Geoffrey} \emph{et~al.}, ``Distilling the knowledge in a neural network,'' \emph{arXiv preprint arXiv:1503.02531}, 2015.

\bibitem{kim2022depthfl}
{Kim, Minjae} \emph{et~al.}, ``Depthfl: Depthwise federated learning for heterogeneous clients,'' in \emph{ICLR}, 2022.

\bibitem{huang2019gpipe}
{Huang, Yanping} \emph{et~al.}, ``Gpipe: Efficient training of giant neural networks using pipeline parallelism,'' \emph{NeurIPS}, vol.~32, 2019.

\bibitem{hubara2017quantized}
{Hubara, Itay} \emph{et~al.}, ``Quantized neural networks: Training neural networks with low precision weights and activations,'' \emph{JMLR}, vol.~18, no.~1, pp. 6869--6898, 2017.

\bibitem{kornblith2019similarity}
{Kornblith, Simon} \emph{et~al.}, ``Similarity of neural network representations revisited,'' in \emph{ICML}.\hskip 1em plus 0.5em minus 0.4em\relax PMLR, 2019, pp. 3519--3529.

\bibitem{chen2022layer}
{Chen, Yixiong} \emph{et~al.}, ``Which layer is learning faster? a systematic exploration of layer-wise convergence rate for deep neural networks,'' in \emph{ICLR}, 2022.

\bibitem{wang2023egeria}
{Wang, Yiding} \emph{et~al.}, ``Egeria: Efficient dnn training with knowledge-guided layer freezing,'' in \emph{EuroSys}, 2023, pp. 851--866.

\bibitem{chen2021communication}
{Chen, Chen} \emph{et~al.}, ``Communication-efficient federated learning with adaptive parameter freezing,'' in \emph{ICDCS}.\hskip 1em plus 0.5em minus 0.4em\relax IEEE, 2021, pp. 1--11.

\bibitem{blondel2008fast}
{Blondel, Vincent D} \emph{et~al.}, ``Fast unfolding of communities in large networks,'' \emph{JSTAT}, vol. 2008, no.~10, p. P10008, 2008.

\bibitem{lai2021oort}
{Lai, Fan} \emph{et~al.}, ``Oort: Efficient federated learning via guided participant selection,'' in \emph{OSDI}, 2021, pp. 19--35.

\bibitem{dl4j}
\BIBentryALTinterwordspacing
``Dl4j,'' 2016. [Online]. Available: \url{https://github.com/deeplearning4j}
\BIBentrySTDinterwordspacing

\bibitem{Monsoon}
\BIBentryALTinterwordspacing
``Monsoon,'' 2023. [Online]. Available: \url{https://www.msoon.com/}
\BIBentrySTDinterwordspacing

\bibitem{htop}
\BIBentryALTinterwordspacing
``Htop,'' 2023. [Online]. Available: \url{https://htop.dev}
\BIBentrySTDinterwordspacing

\bibitem{liu2022no}
{Liu, Ruixuan} \emph{et~al.}, ``No one left behind: Inclusive federated learning over heterogeneous devices,'' in \emph{SIGKDD}, 2022, pp. 3398--3406.

\bibitem{chai2020tifl}
{Chai, Zheng} \emph{et~al.}, ``Tifl: A tier-based federated learning system,'' in \emph{HPDC}, 2020, pp. 125--136.

\bibitem{huang2023elastictrainer}
{Huang, Kai} \emph{et~al.}, ``Elastictrainer: Speeding up on-device training with runtime elastic tensor selection,'' in \emph{MobiSys}, 2023, pp. 56--69.

\bibitem{ma2023cost}
{Ma, Xinyue} \emph{et~al.}, ``Cost-effective on-device continual learning over memory hierarchy with miro,'' in \emph{MobiCom}, 2023, pp. 1--15.

\bibitem{wu2024breaking}
{Wu, Yebo} \emph{et~al.}, ``Breaking the memory wall for heterogeneous federated learning with progressive training,'' \emph{arXiv preprint arXiv:2404.13349}, 2024.

\bibitem{fu2022federated}
{Fu, Xingbo} \emph{et~al.}, ``Federated graph machine learning: A survey of concepts, techniques, and applications,'' \emph{ACM SIGKDD Explorations Newsletter}, vol.~24, no.~2, pp. 32--47, 2022.

\bibitem{li2022one}
{Li, Heju} \emph{et~al.}, ``One bit aggregation for federated edge learning with reconfigurable intelligent surface: Analysis and optimization,'' \emph{TWC}, vol.~22, no.~2, pp. 872--888, 2022.

\bibitem{wang2023fedins2}
{Wang, Jie} \emph{et~al.}, ``Fedins2: A federated-edge-learning-based inertial navigation system with segment fusion,'' \emph{IoTJ}, 2023.

\bibitem{tian2022harmony}
{Tian, Chunlin} \emph{et~al.}, ``Harmony: Heterogeneity-aware hierarchical management for federated learning system,'' in \emph{MICRO}.\hskip 1em plus 0.5em minus 0.4em\relax IEEE, 2022, pp. 631--645.

\end{thebibliography}
\bibliographystyle{IEEEtran}
\end{document}